%% file: main.tex
\documentclass[aps,prd,twocolumn,superscriptaddress,
showpacs,preprintnumbers,nofootinbib,notitlepage]{revtex4-1}

\usepackage{amsmath,amssymb,hyperref,subfigure,xspace}
\usepackage{mciteplus,color,mathtools,graphicx}
\usepackage{braket,bbm,slashed}
\usepackage{multirow}
\usepackage[english]{babel}
\usepackage{mathrsfs}
\usepackage{enumerate}
\usepackage{xspace}
\usepackage{slashed}
\usepackage{wrapfig}
\usepackage[export]{adjustbox}
\usepackage[utf8]{inputenc}

\include{shortcuts}

\hypersetup{ 
    pdfnewwindow=true,      
    colorlinks=true,       
    linkcolor=blue,          
    citecolor=blue,        
    filecolor=blue,      
    urlcolor=blue        
} 

\newcommand{\ceem}{Center for  Exploration  of  Energy  and  Matter, 
Indiana  University,  
Bloomington,  IN  47403,  USA}
\newcommand{\indiana}{Physics  Department,  
Indiana  University,  
Bloomington,  IN  47405,  USA}
\newcommand{\jlab}{Theory Center,
Thomas  Jefferson  National  Accelerator  Facility, 
Newport  News,  VA  23606,  USA}
\newcommand{\icn}{Instituto de Ciencias Nucleares, 
Universidad Nacional Aut\'onoma de M\'exico, Ciudad de M\'exico 04510, Mexico}
\newcommand{\ect}{European Centre for Theoretical Studies in Nuclear Physics and Related areas (ECT$^*$) and Fondazione Bruno Kessler, Villazzano (Trento), I-38123, Italy}
\newcommand{\genova}{INFN Sezione di Genova, Genova, I-16146, Italy}
\newcommand{\cern}{CERN, 1211 Geneva 23, Switzerland}
\newcommand{\ucm}{Departamento de F\'isica Te\'orica, Universidad Complutense de Madrid, 28040 Madrid, Spain}
\newcommand{\seattle}{Physics Department, University of Washington, Seattle, WA 98195-1560, USA}
\newcommand{\jpac}{Joint Physics Analysis Center}

\begin{document}

\title{On the Equivalence of Three-Particle Scattering Formalisms}

\author{A.~W.~Jackura}
\email[email: ]{ajackura@iu.edu}
\affiliation{\indiana}\affiliation{\ceem}

\author{S.~M.~Dawid}
\email[email: ]{sdawid@iu.edu}
\affiliation{\indiana}\affiliation{\ceem}

\author{C.~Fern\'andez-Ram\'irez}
\affiliation{\icn}

\author{V.~Mathieu}
\affiliation{\ucm}

\author{M.~Mikhasenko}
\affiliation{\cern}

\author{A.~Pilloni}
\affiliation{\ect}
\affiliation{\genova}

\author{S.~R.~Sharpe}
\affiliation{\seattle}

\author{A.~P.~Szczepaniak}
\affiliation{\indiana}\affiliation{\ceem}\affiliation{\jlab}

\collaboration{\jpac}

\preprint{JLAB-THY-19-2947}

\begin{abstract}
In recent years, different on-shell 
$\3\to\3$ scattering formalisms have been proposed
to be applied to both lattice QCD and infinite volume scattering processes.
We prove that the formulation in the infinite volume presented
by Hansen and Sharpe in 
Phys.~Rev.~D92, 114509 (2015) and subsequently Brice\~no, Hansen, and Sharpe in
Phys.~Rev.~D95, 074510 (2017) 
can be recovered from the 
$B$-matrix representation, 
derived on the basis of $S$-matrix unitarity, presented by Mai {\em et al.} in Eur.~Phys.~J.~A53, 177 (2017) and
Jackura {\em et al.} in Eur.~Phys.~J.~C79, 56 (2019).
Therefore, both formalisms in the infinite volume 
are equivalent and the physical content is identical. Additionally, the Faddeev equations are recovered in the non-relativistic limit of both representations.
\end{abstract}
\date{\today}
\maketitle

\section{Introduction}\label{sec:intro}
Considerable progress has been achieved recently in determination of the hadron spectrum  from first  principles Quantum Chromodynamics (QCD) \cite{Durr:2008zz,Beane:2011iw,Dudek:2014qha,Williams:2015cvx,Carrillo-Serrano:2015uca,Wilson:2015dqa,Eichmann:2016yit,Eichmann:2018ytt}. Comparison of   experimental data or lattice results with   theoretical models   involves analysis of  partial wave amplitudes in which resonances  appear as pole singularities  in the complex energy and/or angular momentum planes~\cite{Gribov:2009zz}. Thus, a proper description of resonances requires knowledge of  analytic properties of the scattering amplitude. Specifically, the determination of the hadron spectrum from lattice calculations is done using a \textit{quantization condition}~\cite{Luscher:1986pf}, 
which relates discrete energy levels in the finite volume to the infinite-volume, partial waves evaluated at real energy values and later analytically continued to the complex energy plane. 
The quantization  condition has been extensively studied for systems with strong  two-particle  interactions (see, \eg, Ref.~\cite{Briceno:2017max} and references therein). However,  most 
of resonances of current interest decay to three and more particles.

Quantization conditions for three hadrons have been derived 
by various groups using different approaches~\cite{Hansen:2014eka,Hansen:2015zga,Hansen:2016ync,Briceno:2017tce,Briceno:2018mlh,Mai:2017bge,Mai:2018djl,Polejaeva:2012ut,Hammer:2017kms,Hammer:2017uqm,Doring:2018xxx}, as recently reviewed in Ref.~\cite{Hansen:2019nir}. If differences exist between formalisms, this could indicate that important physical content is missing, and that results based on them will lead to unknown systematic errors. Therefore, it is important to unify our understanding of these approaches and establish relationships between all formalisms. 
In addition to quantization conditions, analytic representations of the infinite volume $\3\to\3$ amplitudes are required
to be able to identify 
pole positions of resonances. In this context we discuss two seemingly different approaches and demonstrate their equivalence. 

The first approach is referred to 
as the $B$-matrix representation, and was studied in Refs.~\cite{Mai:2017vot,Jackura:2018xnx}.
Motivated by unitarity of the $S$-matrix, the $B$-matrix refers  to a kernel in a linear integral equation for an elastic $\3\to\3$ connected amplitude.
The $B$-matrix contains both the known long-range  one-particle exchange (OPE) contributions and any short range interactions.
The latter play similar role to 
the $K$-matrix in $\2\to\2$ scattering amplitudes~\cite{Martin:102663}. Aspects of its analytic properties were studied in Ref.~\cite{Jackura:2018xnx}, showing how, besides the unitarity branch point, there are other singularities near the physical region generated by the one-particle exchange, \eg triangle singularities. 
Applying the $B$-matrix formalism in finite volume leads to the quantization condition of Ref.~\cite{Mai:2017bge}. 

The alternative approach was first derived in Ref.~\cite{Hansen:2015zga}, and subsequently generalized to allow for $\2\leftrightarrow \3$ transitions in Ref.~\cite{Briceno:2017tce}.
Hereafter we refer to it as the HS-BHS approach (for the authors initials).
It is a bottom-up construction of the 
$\3\to\3$ amplitude starting from a generic, relativistic effective field theory in finite volume~\cite{Hansen:2014eka}. 
In Ref.~\cite{Hansen:2015zga}, the  corresponding infinite
volume limit of the HS-BHS formalism was derived explicitly, providing an expression for the $\3\to\3$ scattering amplitude
in terms of a $\3\to\3$ analog of the $K$-matrix, referred to here as $\Kc_{\df}$.\footnote{%
This quantity is denoted $\Kc_{\df,3}$ in Ref.~\cite{Hansen:2015zga}.}
This HS-BHS representation is written in terms of two integral equations, 
the first summing one-particle exchanges between $\2\to\2$ subprocesses, and the second involving all orders in $\Kc_{\df}$. 
Since this approach is based on Feynman diagrams, one expects that the result is consistent with unitarity,
and, indeed, very recently this has been shown explicitly~\cite{Briceno:2019muc}.

In the HS-BHS representation, the kernel $\Kc_{\df}$ appears to play a similar role to that of the  short-range part of the $B$-matrix, but it is actually quite different.
It is the main purpose of this work to show  that, nevertheless,  the two representations are equivalent.
Specifically, we derive an integral equation relating the $R$-matrix of Ref.~\cite{Jackura:2018xnx} and the $\3\to\3$ $K$-matrix of Ref.~\cite{Hansen:2015zta}.
Furthermore, we show that the reason for the superficial difference   
lies in the organization of the short-range rescattering effects and difference in the order in which symmetrization of the amplitude is applied.

The paper is organized as follows. Section~\ref{sec:3to3Amps} summarizes 
definitions of on-shell $\3\to\3$ amplitudes and the relevant kinematic variables. Section~\ref{sec:OnShell} reviews  the $B$-matrix and HS-BHS on-shell representations for the $\3\to\3$ amplitude. In Section~\ref{sec:Relation_RK}, we derive the relationship between these two  representations, proving their equivalence. In Section~\ref{sec:Faddeev} we show that in the non-relativistic limit the $B$-matrix can be reduced to the Faddeev equations. 
Our findings  and outlook are summarized in Section~\ref{sec:Conclusion}. We include three technical Appendices. Appendix~\ref{sec:unitarity} reviews the unitarity relation for $\3\to\3$ amplitudes, and  Appendix~\ref{sec:Reexpress} shows how to rewrite the $B$-matrix representation in a form analogous  to that of the HS-BHS representation, which is used in the demonstration of Section~\ref{sec:Relation_RK}. Finally, Appendix~\ref{sec:ImU_proof} proves a crucial relation discussed in  Section~\ref{sec:Relation_RK}.

\section{$\3\to\3$ Amplitudes}\label{sec:3to3Amps}

\begin{figure}[t!]
    \centering
    \includegraphics[ width=0.85\columnwidth]{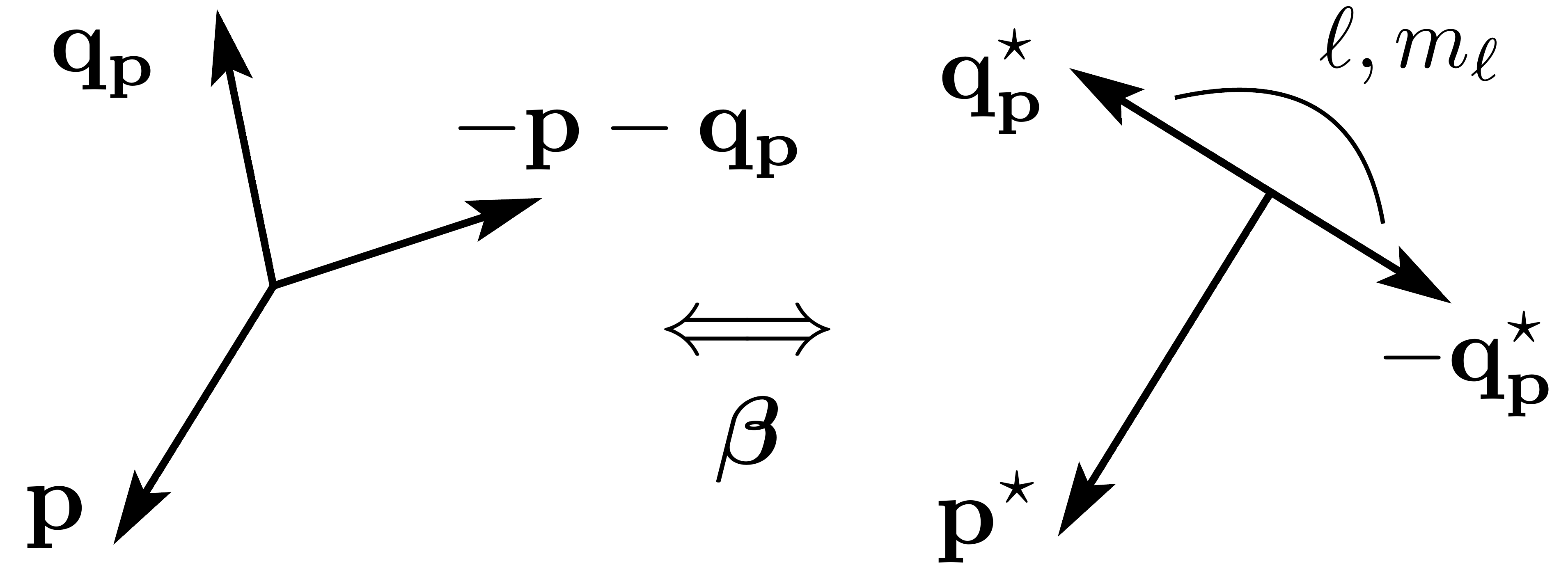}
    \put(-170,0){\colorbox{white}{(a)}}
    \put(-50,0){\colorbox{white}{(b)}}
    \caption{A three-particle state in the (a) CMF with fixed total momentum $\P=\0$ and (b) in the helicity frame at fixed $\P^{\star} - \p^{\star} = \0$. Standard Lorentz transformation with the boost $\bs{\beta} = -(\P - \p) / (E - \omega_{\p})$ transforms the system from the total CMF to the pair rest frame.}
    \label{fig:diag_3to3_CMF_boost}
\end{figure}

We consider the elastic scattering of three spinless identical particles of mass $m$, 
\eg, $3\pi^+\to 3\pi^+$ scattering. Note that Ref.~\cite{Jackura:2018xnx} considered distinguishable particles, while here we consider identical particles to compare with Refs.~\cite{Hansen:2015zga,Briceno:2017tce}. Internal symmetries such as isospin are not considered, but can be included in a straightforward manner. The initial and final three-particle state have a total energy momentum $P=(E,\P)$ and $P'=(E',\P')$, respectively.
This exemplifies a convention we use throughout, namely that primed (unprimed) variables denote quantities in the final (initial) state.
Total energy-momentum is conserved, as is the three-particle invariant mass squared
\beq
s \equiv P^2 = E^2 - \P^2,
\eeq
which lies in the range $(3m)^2 \le s < s_{\textrm{inel}}$, where $s_{\textrm{inel}}$ the first inelastic threshold. 
It is convenient to split the three-particle kinematics into a \textit{pair}\footnote{%
Other commonly used terms for the pairs found in the literature are \textit{dimers} and \textit{isobars}. More precisely, however, isobars refer to partial wave amplitudes of the $\2\to\2$ subsystem in a definite partial wave with only the unitarity branch cut \cite{Jackura:2018xnx}. We will refrain from this terminology in an attempt to unify different approaches and avoid confusion.}
and a \textit{spectator}. The spectator is a single particle that has  energy-momentum $p = (\omega_{\p},\p)$, where $\omega_{\p} = \sqrt{m^2 + \p^2}$ is the on-shell energy. 
The pair then consists of the other two particles with total energy-momentum given by 
\beq
P_{\p} \equiv P - p = (E_{\p},\P_{\p}) = (E-\omega_{\p},\P - \p),
\eeq
where the subscript labels the spectator associated with the pair. The invariant mass squared of the pair is 
\beq
\sigma_{\p} \equiv P_{\p}^2 = (P - p)^2 = (E - \omega_{\p})^2 - (\P - \p)^2,
\eeq
for which the physical region is $(2m)^2 \le \sigma_{\p} \le (\sqrt{s} - m)^2$.
In the helicity frame of the two particle subsystem, \ie, the pair rest frame where $\P^{\star} - \p^{\star} = \0$, the three momenta of particles inside the pair are  $\q_{\p}^{\star}$ and  $-\q_{\p}^{\star}$, respectively.
In this frame, the spectator defines the $-z$-axis, and the $y$-axis is perpendicular to the plane formed by the three particles.
Kinematic quantities without a $\star$ are taken to be in the total center of momentum frame (CMF), \ie where $\P = \0$. Figure~\ref{fig:diag_3to3_CMF_boost} illustrates the momenta of the three particles in these frames. The magnitude of the spectator momentum in the CMF is given by
\beq\label{eq:spectator_momentum}
p \equiv \lvert \p \rvert = \frac{1}{2s} \lambda^{1/2}(s,m^2,\sigma_{\p}),
\eeq
where $\lambda(x,y,z) = x^2 + y^2 + z^2 - 2(xy + yz + zx)$ is the K\"all\'en triangle function, while the relative momentum of the pair in the helicity frame is given by
\begin{equation}
q_{\p}^{\star}  = \frac{1}{2\sqrt{\sigma_{\p}}} \, \lambda^{1/2}(\sigma_{\p},m^2,m^2) = \frac{1}{2}\sqrt{\sigma_{\p} - 4m^2}.
\end{equation}
The final state variables have similar expressions with $\p$ replaced by $\p'$.

The elastic $\3\to\3$ scattering amplitude, $\Mc$, is a Lorentz scalar that depends on eight kinematic variables, and is defined via 
\beq\label{eq:Smatrix}
\bra{\textrm{out}} T \ket{\textrm{in}} = (2\pi)^{4} \delta^{(4)}(P' - P) \,  \Mc,
\eeq
where the $T$-matrix is as usual given by $S = \mathbbm{1} + i T$. 
The Dirac $\delta$-function ensures total energy-momentum conservation. 
Due to Bose statistics, the amplitude is symmetric under interchange of any pair of particles in the initial or final state. In the following, we express $\Mc$ in terms of an unsymmetrized amplitude  $[\,\Mc_{\p'\p}\,]_{\ell' m_{\ell}' ; \ell m_{\ell} }$, which is expressed in the mixed $p\ell m_{\ell}$-basis, \ie,  it  depends on the spectator momenta, $\p$, and the angular momentum of the pair, $(\ell,m_{\ell})$. The fully symmetric amplitude is then obtained by replacing the dependence on spin by that of
the corresponding spherical angles, (through
multiplication by spherical harmonics),
and by symmetrizing
with respect to particle permutations, an operation
that we denote by $\Sc$,
\beq\label{eq:SymmAmp}
\Mc = \Sc \left\{ 4\pi  \sum_{ \substack{ \ell',m_{\ell}' \\ \ell,m_{\ell}} } Y^*_{\ell' m_{\ell}'}({\bh{\q}}_{\p'}^{\star}) \left[ \, \Mc_{\p'\p} \, \right]_{\ell' m_{\ell}' ; \ell m_{\ell}} Y_{\ell m_{\ell}}({\bh{\q}}_{\p}^{\star}) \right\}.
\eeq
The unsymmetrized amplitudes are infinite dimensional matrices in  the $\ell m_{\ell}$-space. Note that since the particles are identical, due to Bose symmetry, 
all odd-$\ell$ amplitudes must be zero. We will often leave the indices implicit and consider amplitudes as matrices in the $\ell m_{\ell}$-space. 
The $\p$-dependence will be left explicit unless otherwise noted. The isobar representation of Ref.~\cite{Jackura:2018xnx} is identical to the symmetrization operation in Eq.~\eqref{eq:SymmAmp}. 
However, unlike in Ref.~\cite{Jackura:2018xnx}, here we do not truncate the spin of the pair to some maximum value, and instead work formally with infinite-dimensional matrices. In practice, one must truncate the partial waves, in which the resummation strategy presented in Eq.~(17) of Ref.~\cite{Jackura:2018xnx} can be used to recover cross channel effects.

The scattering amplitude $\Mc$ contains disconnected and connected contributions.
The disconnected terms in $\Mc$, denoted hereafter by $\Fc$,\footnote{%
For convenience, we collect here the correspondences
between our amplitude definitions and those of Ref.~\cite{Hansen:2015zga}. 
The $\2\to\2$ amplitude is  $\Fc=\Mc_{2}$({\textrm{Ref.}\cite{Hansen:2015zga}}),
the connected $\3\to\3$ amplitude is $\Ac=\Mc_{3}^{(u,u)}$({\textrm{Ref.}\cite{Hansen:2015zga}}), and 
(as already noted above) 
$\Kc_{\df}=\Kc_{\df,3}$({\textrm{Ref.}\cite{Hansen:2015zga}}).
In addition, the ladder series encountered below is
$\Dc = \Dc^{(u,u)}$({\textrm{Ref.}\cite{Hansen:2015zga}}).}
are associated with  $\2\to\2$ process in which the spectator particle does not participate   \beq\label{eq:3to3_split}
\Mc_{\p'\p} = \delta_{\p'\p} \, \Fc_{\p} + \Ac_{\p'\p}.
\eeq
with the spectator momentum conserving  
$\delta$-function, $\delta_{\p'\p} \equiv (2\pi)^{3} \, 2\omega_{\p} \, \delta^{(3)}(\p'-\p)$, written explicitly in front of $\Fc_{\p}$.  
The amplitude $\Fc_{\p}$ is diagonal in the spin variables, and depends solely on the single, scalar variable, the pair's invariant mass,  $\sigma_{\p} = \sigma_{\p'}$, 
\begin{align}
\left[ \, \Fc_{\p} \, \right]_{\ell' m_{\ell}' ; \ell m_{\ell}}  & = \delta_{\ell' \ell} \delta_{m_{\ell}' m_{\ell}}  \, \Fc_{\ell} (\sigma_{\p}), \nn \\
\
& = \includegraphics[width=0.25\textwidth, valign=c]{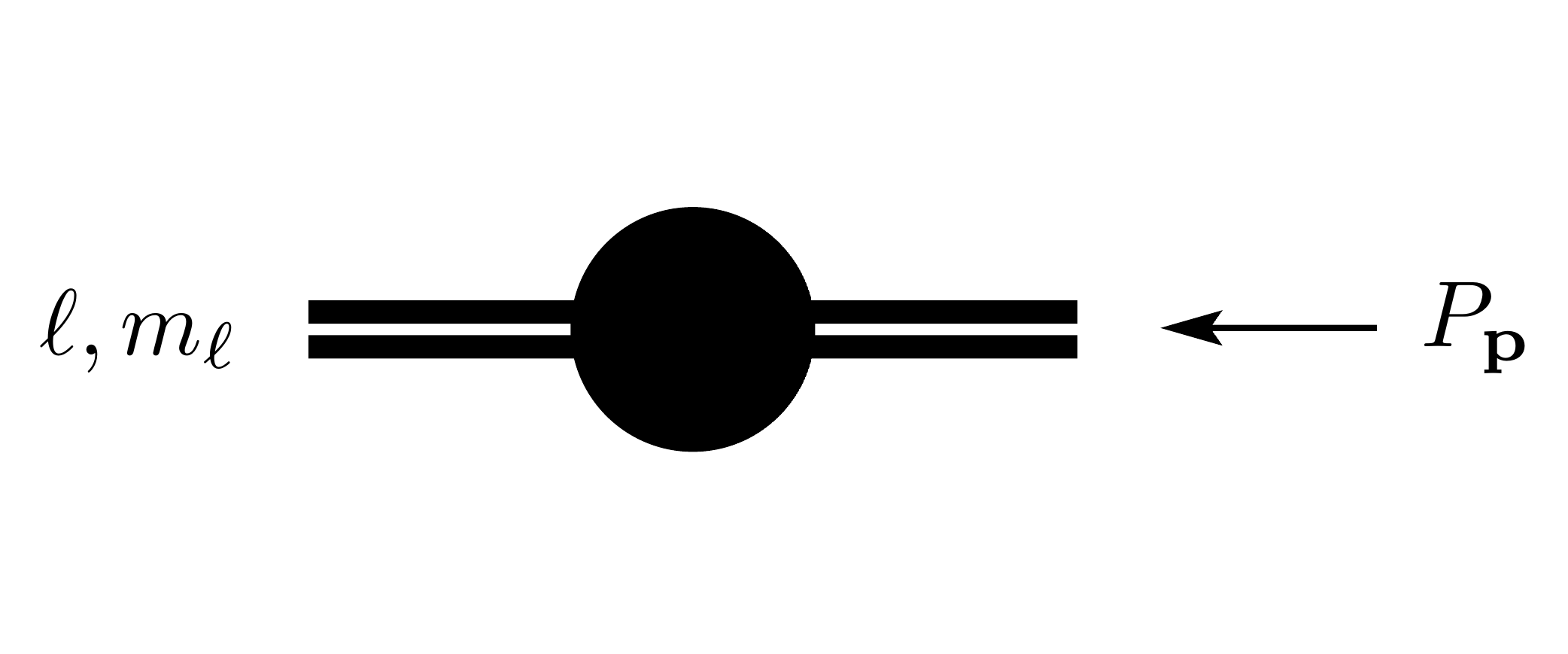} 
\end{align}
The second term,  $\Ac_{\p'\p}$  in Eq.~\ref{eq:3to3_split} is the connected $\3\to\3$ amplitude and it contains offdiagonal contributions in spin indices, 
\begin{align}
\left[ \, \Ac_{\p'\p} \, \right]_{\ell' m_{\ell}' ; \ell m_{\ell}} & = \Ac_{\ell' m_{\ell}' ; \ell m_{\ell}} (\p',s,\p). \nn \\[8pt]
\
& = \includegraphics[width=0.25\textwidth, valign=c]{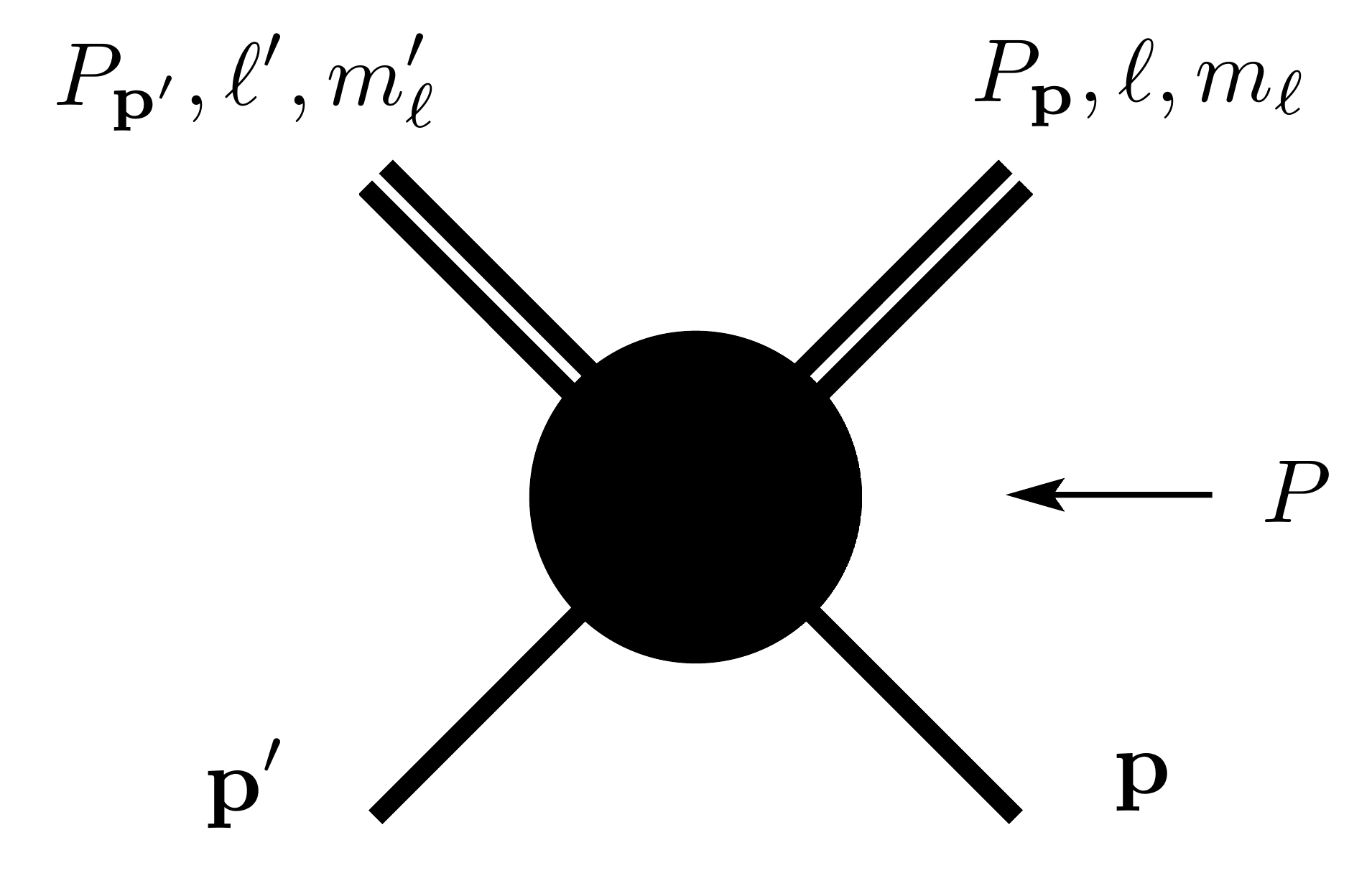}
\end{align}
In addition to its implicit dependence on $\ell$, $m_\ell$, $\ell'$ and $m'_\ell$, $\Ac_{\p'\p}$  depends on the remaining four independent
variables. 
Convenient choices are $s$, the initial and final pair invariant mass squares, $\sigma_{\p}$ and $\sigma_{\p'}$, and the scattering angle between spectators in the CMF, $\Theta_{\p'\p}$, defined as $\cos\Theta_{\p'\p} \equiv  {\bh{\p}}'\cdot{\bh{\p}} = {\bh{\P}}_{\p'}\cdot{\bh{\P}}_{\p}$.

\section{On-Shell Representations}\label{sec:OnShell}
Our interest is in constructing on-shell representations for the connected $\3\to\3$ scattering amplitude. Here we review the relevant features of the $B$-matrix representation discussed in Ref.~\cite{Jackura:2018xnx} and the HS-BHS representation of Ref.~\cite{Hansen:2015zga}. 

\subsection{B-Matrix Representation}\label{sec:Bmatrix}
\begin{figure}[t!]
    \centering
    \includegraphics[ width=0.99\columnwidth]{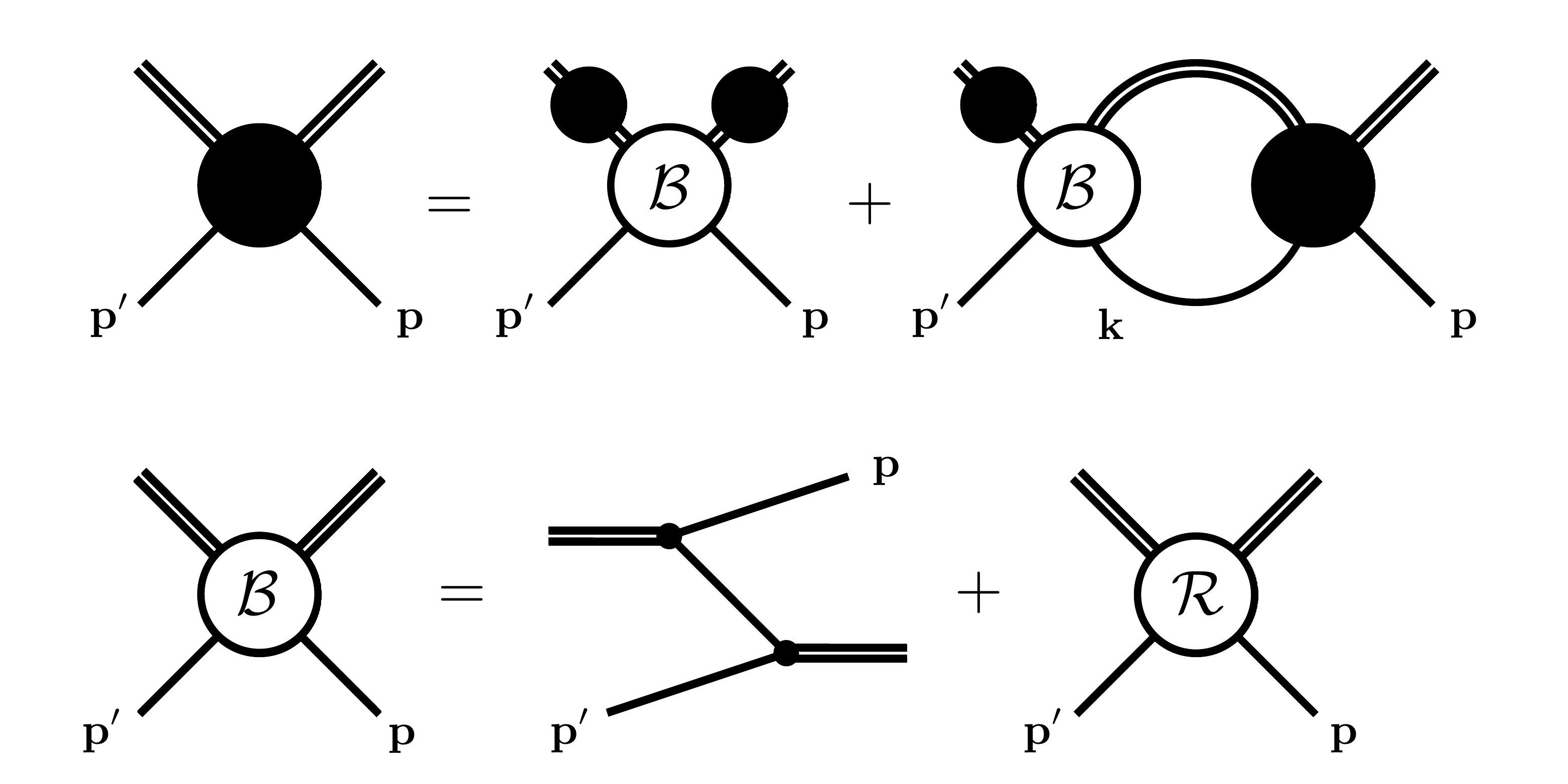}
    \put(-245,90){\colorbox{white}{(a)}}
    \put(-245,25){\colorbox{white}{(b)}}
    \caption{Diagrammatic representation of (a) the $B$-matrix representation for the on-shell amplitude, Eq.~\eqref{eq:Bmatrix}, and (b) the $B$-matrix which is composed of the OPE $\Gc_{\p'\p}$, Eq.~\eqref{eq:OPE}, and the $R$-matrix.}
    \label{fig:diag_3to3_B-matrix_amplitude}
\end{figure}

As discussed in Refs.~\cite{Mai:2017vot,Jackura:2018xnx},
the $B$-matrix is an on-shell representation for the connected $\3\to\3$ amplitude that was constructed to satisfy elastic $\3\to\3$ unitarity. In the $p\ell m_{\ell}$-basis, the $B$-matrix representation it leads to the integral equation
\beq\label{eq:Bmatrix}
\Ac_{\p'\p} = \Fc_{\p'} \, \Bc_{\p'\p} \, \Fc_{\p} + \int_{\k} \Fc_{\p'} \, \Bc_{\p'\k} \, \Ac_{\k\p}
\eeq
where $\Bc_{\p'\p} = \Gc_{\p'\p} + \Rc_{\p'\p}$ is the $B$-matrix driving term, with $\Gc_{\p'\p}$ being the OPE contribution\footnote{%
In Ref.~\cite{Jackura:2018xnx} we denoted the OPE contribution by the symbol $\Ec$, while here we use $\Gc$ to provide a closer connection to the notation
of Ref.~\cite{Hansen:2015zga}.}
and $\Rc_{\p'\p}$ a real function called the $R$-matrix. Figure~\ref{fig:diag_3to3_B-matrix_amplitude} shows a diagrammatic representation of Eq.~\eqref{eq:Bmatrix}. By construction, Eq.~\eqref{eq:Bmatrix} satisfies the $\3\to\3$ unitarity relation given that $\Fc_{\p}$ is known, as demonstrated in Appendix~\ref{sec:unitarity}. Equation~\eqref{eq:Bmatrix} is an infinite dimensional matrix equation in $(\ell,m_{\ell})$-space, and the integration over the spectator momenta includes the measure,
\beq
\int_{\k} \equiv \int \frac{\diff^{3} \k}{(2\pi)^{3} \, 2\omega_{\k}}.
\eeq
The integration ranges over all momenta, or equivalently in $-\infty \le \sigma_{\k} \le (\sqrt{s}-m)^2$ and over the entire solid angle of the spectator. The $|{\bf k}| \to \infty$ ($\sigma_{\bf k} \to -\infty$) limit is divergent and needs to be regulated. The preferred  option is to restrict integration region 
to $4m^2 \le \sigma_{\k} \le (\sqrt{s} - m)^2$, which is the only domain of $\sigma_{\bf k}$ that is actually restricted by $\3\to\3$ unitarity~\cite{Jackura:2018xnx}. Beyond this region, one deals with unphysical (off-mass shell)  amplitudes, which depend on unknown parameters, \eg subtraction constants. 

The OPE amplitude is given by 
\begin{align}\label{eq:OPE}
\left[ \,\Gc_{\p'\p} \, \right]_{\ell' m_{\ell}' ; \ell m_{\ell}} & = \left( \frac{p_{\p'}^{\star}}{q_{\p'}^{\star}} \right)^{\ell'} \frac{4\pi \, Y_{\ell' m_{\ell}'}({\bh{\p}}_{\p'}^{\star}) Y^{*}_{\ell m_{\ell}}({\bh{\p}}_{\p}'^{\star}) }{m^2 - (P_{\p} - p')^2 - i\epsilon} \left( \frac{p_{\p}'^{\star}}{q_{\p}^{\star}} \right)^{\ell} , \nn \\[8pt]
\
& = \includegraphics[width=0.25\textwidth, valign=c]{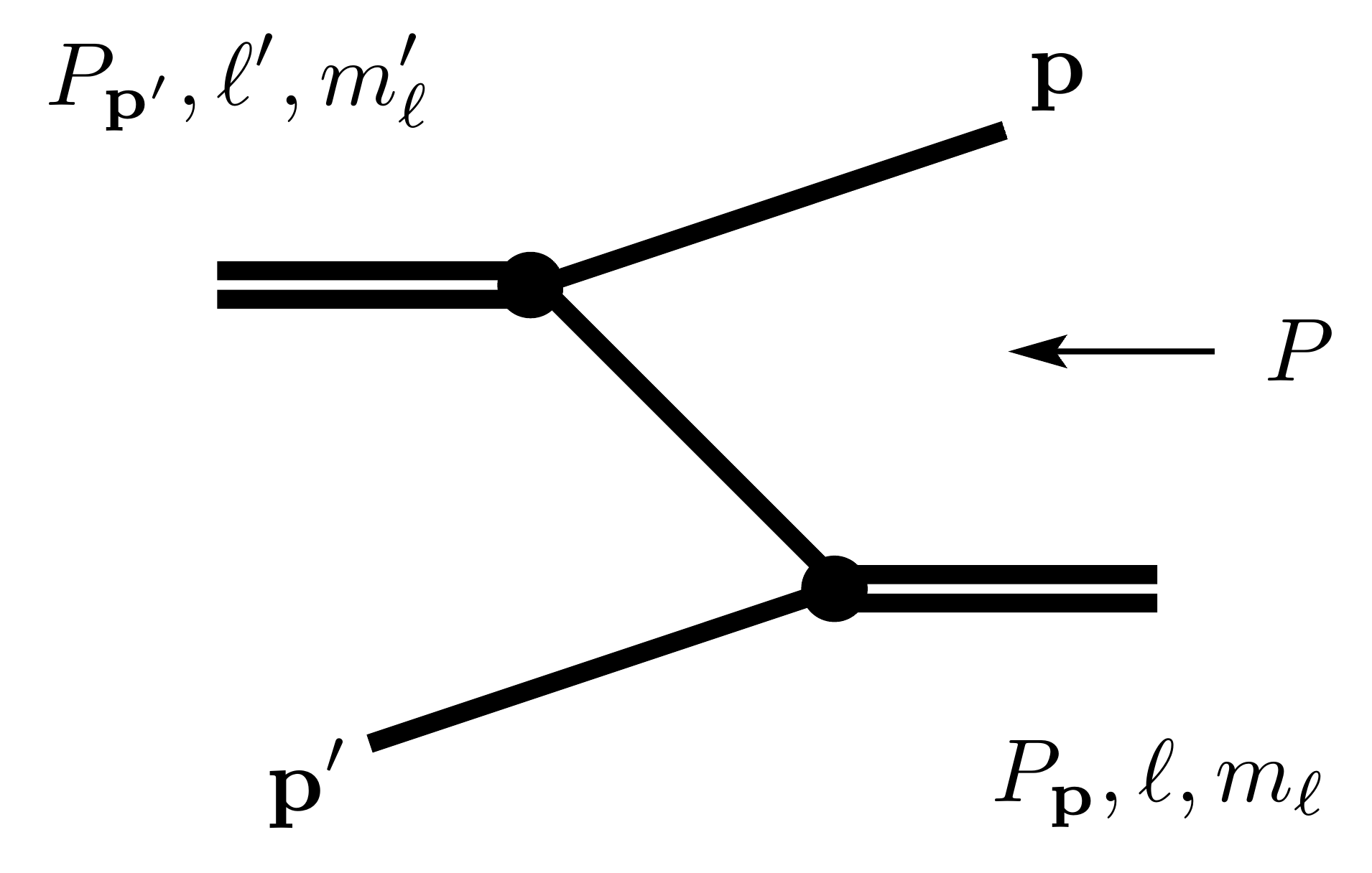}
\end{align}
where ${\bh{\p}}_{\p'}^{\star}$ is the direction of momentum of the initial state spectator in the final state pair rest frame. Similarly, ${\bh{\p}}_{\p}'^{\star}$ is the orientation of the final state spectator in the initial state pair rest frame. 
The magnitude of these momenta are  
\begin{align}\label{eq:spectator_cross}
p_{\p'}^{\star} & = \frac{1}{2\sqrt{\sigma_{\p'}}} \, \lambda^{1/2}\left((P_{\p'} - p )^2,\sigma_{\p'},m^2 \right), \nn \\
\
p_{\p}'^{\star} & = \frac{1}{2\sqrt{\sigma_{\p}}} \, \lambda^{1/2}\left((P_\p - p' )^2,\sigma_{\p},m^2 \right).
\end{align}
Note that 
energy-momentum conservation gives  $P_{\p} - p' = P_{\p'} - p$.  
The normalization of the barrier factors is chosen such that they are equal to one when the  exchanged particle is on its mass shell, $(P_{\p} - p')^2 = m^2$. 

Our definition of $\Gc$ differs from the corresponding quantity in Ref.~\cite{Hansen:2015zga}, denoted $\Gc^\infty$,
in three ways.  First, there is a difference in overall sign. 
We find the choice in Eq.~\eqref{eq:OPE} more convenient since it has a positive imaginary part, which avoids several
minus signs in expressions.
Second, $\Gc^\infty$ contains a cutoff function, which serves to cut off the integrals over spectator momenta, which in Ref.~\cite{Hansen:2015zga} run over all values.
Third, the form given in Ref.~\cite{Hansen:2015zga} has the nonrelativistic 
form of the pole in the denominator, in contrast to the relativistic form
in Eq.~\eqref{eq:OPE}. However, in recent applications of the BH+BHS formalism,
e.g., in Refs.~\cite{Briceno:2017tce,Blanton:2019igq},
the relativistic form is used.
We also note that the barrier factors in $\Gc$ are not required from unitarity, but are included so as to match those in $\Gc^\infty$,
where these factors are included since they are needed in the finite-volume analysis.

\begin{figure*}[t!]
    \centering
    \includegraphics[ width=0.8\textwidth]{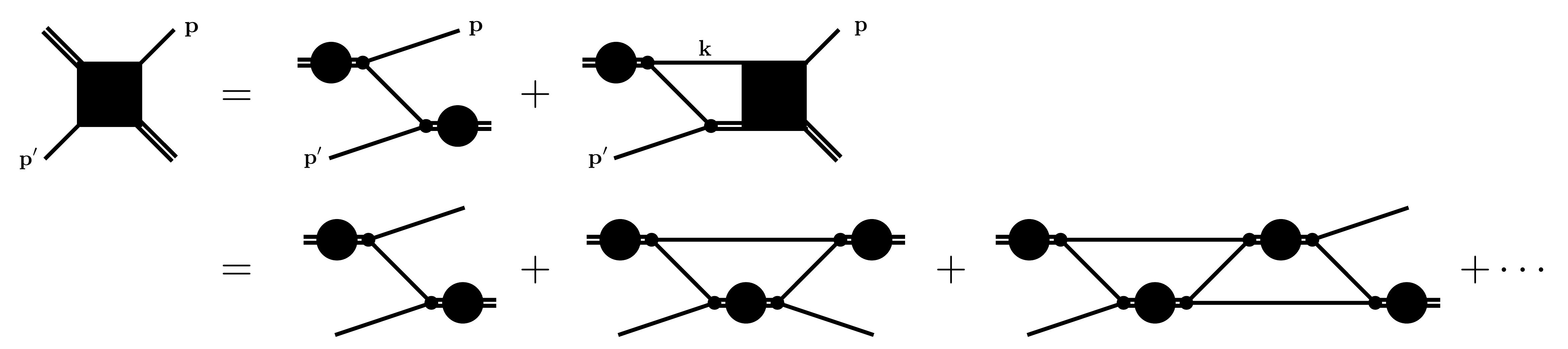}
    \caption{Diagrammatic representation for the ladder series generated by particle exchanges, Eq.~\eqref{eq:OPE_ladder}, where the black box represents $\Dc_{\p'\p}$.}
    \label{fig:diag_3to3_ladder}
\end{figure*}

As with the $K$-matrix representation for $\2\to\2$ scattering (see Appendix~\ref{sec:unitarity}), the $R$-matrix is a real function that represents the dynamical content of the three-particle system, \eg, the short range forces between pions in elastic $3\pi$ scattering. Included in this term are virtual exchange processes giving rise to left hand cuts, and higher multiparticle thresholds, \eg, $5\pi$ production, which are off-shell in the kinematic domain of elastic $3\pi$ scattering. In principle, given a specific theory, $\Rc_{\p'\p}$ can be computed. Alternatively, given some data, \eg, from lattice QCD calculations, $\Rc_{\p'\p}$ can be determined via the quantization condition of Ref.~\cite{Mai:2017bge}.
In the limit where there are no short-range three-body interactions, \ie, when $\Rc_{\p'\p} = 0$, Eq.~\eqref{eq:Bmatrix} reduces to a solution composed of entirely exchanges between $\2\to\2$ subprocesses. We denote this solution as $\Dc_{\p'\p}$, which is called the \textit{ladder series} and is the solution of the integral equation, 
\beq\label{eq:OPE_ladder}
\Dc_{\p'\p} = \Fc_{\p'} \, \Gc_{\p'\p} \, \Fc_{\p} + \int_{\k} \Fc_{\p'} \, \Gc_{\p'\k} \, \Dc_{\k\p}.
\eeq
Although the ladder series is an explicit solution of the $\3\to\3$ unitarity relations, it is dynamically controlled by long range exchanges between $\2\to\2$ subsystems. Once the $\2\to\2$ amplitudes are known, then solutions of Eq.~\eqref{eq:OPE_ladder} are completely fixed. Figure~\ref{fig:diag_3to3_ladder} shows a diagrammatic representation of the ladder series solution.

As noted in Ref.~\cite{Jackura:2018xnx}, and further explored in Ref.~\cite{Mikhasenko:2019vhk},
(see also Appendix~\ref{sec:Reexpress}), the $B$-matrix representation can be rewritten into a form where the ladder series is explicitly separated from the remaining three-particle interaction. This separation is useful in comparing to the HS-BHS equations as they schematically follow the same procedure. Genuine three-body interactions are introduced through an additional term to the ladder series solution, known in Ref.~\cite{Hansen:2015zga} as \textit{divergence-free} amplitudes.\footnote{%
Divergence-free in that the kinematic singularities from all long range exchanges are not included, as they are contained in $\Dc_{\p'\p}$ in Eq.~\eqref{eq:OPE_ladder}.}
Following the derivation in Appendix~\ref{sec:Reexpress}, the resulting $\3\to\3$ amplitude has the form
\beq\label{eq:Bmat_convert}
\Ac_{\p'\p} = \Dc_{\p'\p} + \int_{\k'}\int_{\k} \wt{\Lc}_{\p'\k'} \, \wt{\Tc}_{\k'\k} \, \wt{\Lc}_{\k\p},
\eeq
where the first term is the ladder series which satisfies Eq.~\eqref{eq:OPE_ladder} and the second term is the divergence-free amplitude. The second term contains the amputated $\wt{\Tc}_{\p'\p}$ amplitude, which is determined by the integral equation
\beq\label{eq:Bmat_T}
\wt{\Tc}_{\p'\p} = \Rc_{\p'\p} + \int_{\k'}\int_{\k} \Rc_{\p'\k} \, \wt{\Lc}_{\k'\k} \, \wt{\Tc}_{\k\p},
\eeq
as well as a $\2\to\2$ rescattering function, $\wt{\Lc}_{\p'\p}$,
\begin{align}
\wt{\Lc}_{\p'\p} & = \Fc_{\p}\,\delta_{\p'\p} + \Dc_{\p'\p}  \\
\
& = \Fc_{\p} \, \delta_{\p'\p} + \int_{\k} \Fc_{\p'} \,\Gc_{\p'\k}\,\wt{\Lc}_{\k\p}.
\label{eq:Bmat_L}
\end{align}
In the second line we used the fact that $\Dc_{\p'\p}$ satisfies Eq.~\eqref{eq:OPE_ladder} to write the rescattering dressing function as an integral equation. Tildes on $\wt{\Tc}_{\p'\p}$ and $\wt{\Lc}_{\p'\p}$ are used to distinguish these quantities from the corresponding HS-BHS amplitudes, 
which, though similar, have different definitions. We will discuss these differences later, when we perform the direct comparison. 
The interpretation of the divergence-free amplitude is now straightforward: $\wt{\Tc}_{\p'\p}$ is an amplitude that involves $\3\to\3$ interactions via short-range dynamics. Rescattering functions then dress the initial and final states with all rescatterings that involve $\2\to\2$ processes, \ie, with direct $\2\to\2$ amplitudes or exchanges in the ladder series. The original $B$-matrix, Eq.~\eqref{eq:Bmatrix}, explicitly shows only the direct $\2\to\2$ amplitudes dressing the initial and final state, while the ladder series remains hidden. 

In Ref.~\cite{Mikhasenko:2019vhk}, the
authors proposed an initial-final-state factorization model of the short-range amplitude, $\Rc_{\p'\p}$, under which Eq.~\eqref{eq:Bmat_T} becomes algebraic, with the ladder series having a different definition for the real part.
This construction may prove practical for analysis of data 
relevant to resonance phenomena.

\subsection{HS-BHS Representation}\label{sec:BHS}
We now turn to the definitions of the on-shell $\3\to\3$ scattering equations of HS-BHS as given in Ref.~\cite{Hansen:2015zga}. We remind the reader that the unsymmetrized elastic $\3\to\3$ amplitude, $\Ac_{\p'\p}$, is a matrix in the angular momentum space of the pair labeled by the spectator. The unsymmetrized elastic $\3\to\3$ amplitude in HS-BHS representation is given via the integral equation
\beq\label{eq:BHS_3to3Amp}
\Ac_{\p'\p} = \Dc_{\p'\p} + \int_{\k'}\int_{\k} \Lc_{\p'\k'} \, \Tc(\k',\k) \, \Lc_{\k'\p}^{\bs{\top}}.
\eeq
The symmetrized amplitude can be recovered as in Eq.~\eqref{eq:SymmAmp}. The ladder series, $\Dc_{\p'\p}$, is defined exactly like in Eq.~\eqref{eq:OPE_ladder}, and the end cap operators, $\Lc_{\p'\p}$
are defined by\footnote{%
$\Lc$ is the same as the quantity $\Lc^{(u,u)}$ of Ref.~\cite{Hansen:2015zga}. To see this requires accounting for the different integration measures used in the two works: our measure includes a factor of $1/(2\omega_{\p})$ that is not present in the measure of Ref.~\cite{Hansen:2015zga}.}
\beq\label{eq:BHS_Ldef}
\Lc_{\p'\p} = \left( \frac{1}{3} + \Fc_{\p'} \, i\rho_{\p'}  \right) \delta_{\p'\p} + \Dc_{\p'\p} \,i\rho_{\p},
\eeq
with $\Lc_{\p'\p}^{\top}$ defined with $i\rho_{\p}$ on the left of $\Fc_{\p'}$ and $\Dc_{\p'\p}$.
The quantity $\rho_{\p}$ is the phase space factor for the two
particles in the pair,\footnote{%
In Ref.~\cite{Hansen:2015zga}, 
the two-body phase space is defined slightly differently,
$\rho_{\p}({\textrm{Ref.\cite{Hansen:2015zga}}}) 
= -i \rho_{\p} H(\p)$. Thus there are no explicit factors of
$i$ in the expression for $\Lc$ in Ref.~\cite{Hansen:2015zga},
whereas we prefer here to keep such factors explicit.
The object $H(\p)$ is a cutoff function, absent here because our
momentum integrals implicitly include an ultraviolet cutoff, as
discussed above.}
\beq\label{eq:2bodyPS}
\left[ \, \rho_{\p} \, \right]_{\ell' m_{\ell}' ; \ell m_{\ell}} = \delta_{\ell'\ell}  \delta_{m_{\ell}' m_{\ell}} \frac{1}{2!} \frac{1}{16\pi} \sqrt{1 - \frac{4m^2}{\sigma_{\p}}} \, 
\eeq
where the $2!$ is the symmetry factor. Finally, $\Tc(\p',\p)$ is defined via the integral equation
\begin{align}
\Tc(\p',\p) & = \Kc_{\df}(\p',\p) \nn \\
\
& + \int_{\k'}\int_{\k}\Kc_{\df}(\p',\k') \, i\rho_{\k'} \, \Lc_{\k'\k} \, \Tc(\k,\p),
\end{align}
where $\Kc_{\df}(\p',\p)$ is the three-particle $K$-matrix. 
The amplitudes $\Tc(\p',\p)$ and $\Kc_{\df}(\p',\p)$ are matrices in angular momenta,
\begin{align}
\left[ \, \Kc_{\df}(\p',\p) \, \right]_{\ell'm_{\ell}';\ell m_{\ell}} & = \Kc_{\df,\ell'm_{\ell}';\ell m_{\ell}}(\p',s,\p),
\end{align}
and similarly for $\Tc(\p',\p)$, however we denote them differently than all other amplitudes due to their symmetry properties. As defined in Ref.~\cite{Hansen:2015zga}, $\Tc(\p',\p)$ and $\Kc_{\df}(\p',\p)$ are symmetric under interchange of any pair of initial or final state particles, after we sum over the product of the amplitude and its spherical harmonics of the pair orientations. Thus, the symmetric divergence-free $K$ matrix is given by
\beq\label{eq:SymmK}
\Kc_{\df} = 4\pi \sum_{\substack{ \ell',m_{\ell}' \\ \ell,m_{\ell} } }  Y^{*}_{\ell' m_{\ell}'}({\bh{\q}}_{\p'}^{\star}) \, \Kc_{\df,\ell' m_{\ell}' ; \ell m_{\ell}}(\p',s,\p) \, Y_{\ell m_{\ell}}({\bh{\q}}_{\p}^{\star}) ,
\eeq
with a similar expression for $\Tc$. Note that Eq.~\eqref{eq:SymmK} is different from Eq.~\eqref{eq:SymmAmp} since the latter requires a further symmetrization operation. The $K$ matrix on the left hand side in Eq.~\eqref{eq:SymmK} is fully symmetric under interchange of any pair of particles in either the initial or final state.

$\Tc(\p',\p)$ is viewed as an amputated amplitude
for which, in addition to all $\2\to\2$ rescatterings being removed, the possibility of no rescattering in either initial or final is included. 
This possibility is allowed by the term involving the constant $1 / 3$ in Eq.~\eqref{eq:BHS_Ldef}. 
The factor of $1/3$ is due to the partial wave definitions of $\Tc(\p',\p)$ and $\Kc_{\df}(\p',\p)$, Eq.~\eqref{eq:SymmK}, which is different than Eq.~\eqref{eq:SymmAmp}. 
Therefore, when we symmetrize the amplitude in Eq.~\eqref{eq:BHS_3to3Amp}, 
we would overcount the terms with 
no rescatterings if the $1/3$ were not present.

\section{Equivalence of the $B$-matrix and HS-BHS Representations}\label{sec:Relation_RK}

Having established the $B$-matrix and HS-BHS equations, we now show that they are equivalent. To do so we assume that the $\3\to\3$ amplitudes in both representations are equal, 
and search for a relation between $\Rc$ and $\Kc_{\df}$.
We first express the HS-BHS end caps, $\Lc_{\p'\p}$, in terms of the $B$-matrix rescattering functions, $\wt{\Lc}_{\p'\p}$,
\beq\label{eq:L_relate}
\Lc_{\p'\p} = \frac{1}{3}\delta_{\p'\p} + \wt{\Lc}_{\p'\p} \, i\rho_{\p}.
\eeq
The result for $\Lc_{\p'\p}^{\top}$ simply has 
$i\rho_{\p'}$ and $\wt{\Lc}_{\p'\p}$ interchanged.
We will find that the first term of Eq.~\eqref{eq:L_relate} 
can be traced to the differences in symmetrization and removal of $\2\to\2$ rescatterings between $\wt{\Tc}_{\p'\p}$ and $\Tc(\p',\p)$, while the $i\rho_{\p}$ factor in the second term is due to a difference in the definition of on-shell amputation.

To proceed, we rewrite Eq.~\eqref{eq:L_relate} as
\beq\label{eq:LU}
\Lc_{\p'\p} = \int_{\k} \wt{\Lc}_{\p'\k} \, \Uc_{\k\p},
\eeq
where $\Uc_{\p}$ is the ``conversion factor''
\begin{align}\label{eq:Ufactor}
\Uc_{\p'\p} & = i\rho_{\p'} \, \delta_{\p'\p} + \frac{1}{3} \wt{\Lc}_{\p'\p}^{-1}  \\
\
& = i\rho_{\p'} \, \delta_{\p'\p} + \frac{1}{3} \Fc_{\p'}^{-1} \delta_{\p'\p} - \frac{1}{3} \Gc_{\p'\p},
\end{align}
where the second line follows from the inverse of $\wt{\Lc}_{\p'\p}$ obtained from Eq.~\eqref{eq:Bmat_L}.
In a similar manner, the transpose is given by
$\Lc_{\p'\p}^{\top} = \int_{\k} \Uc_{\p'\k} \, \wt{\Lc}_{\k\p}$.
Now, equating the expressions for $\Ac$ in the two formalisms,
Eqs.~\eqref{eq:Bmat_convert} and \eqref{eq:BHS_3to3Amp}, 
and using Eq.~\eqref{eq:LU}, we find the equivalence if the
following relation holds,
\beq\label{eq:T_relate}
\wt{\Tc}_{\p'\p} = \int_{\k'}\int_{\k} \Uc_{\p'\k'} \, \Tc(\k',\k) \, \Uc_{\k,\p}.
\eeq
The amplitudes $\wt{\Tc}_{\p'\p}$ and $\Tc(\p',\p)$ can be formally solved in terms of $\Rc_{\p'\p}$ and $\Kc_{\df}(\p',\p)$, respectively, as one does in matrix equations, \eg $\wt{\Tc} = \left[ \, 1 - \Rc \, \wt{\Lc} \, \right]^{-1} \Rc$, which is a matrix in both angular and spectator momenta. Combining the formal solutions for $\wt{\Tc}_{\p'\p}$ and $\Tc(\p',\p)$, the relation Eq.~\eqref{eq:T_relate}, and using the definition of $\Uc_{\p'\p}$ in Eq.~\eqref{eq:Ufactor}, we arrive at an integral equation relating $\Rc_{\p'\p}$ and $\Kc_{\df}(\p',\p)$
\begin{align}\label{eq:IntegralRK}
\Rc_{\p'\p} & = \int_{\k'}\int_{\k} \Uc_{\p'\k'} \, \Kc_{\df}(\k',\k) \,\Uc_{\k\p}  \nn \\ 
\
& - \frac{1}{3} \int_{\k'}  \int_{\k} \Uc_{\p'\k'} \, \Kc_{\df}(\k',\k) \, \Rc_{\k\p}
\end{align}
If this equation holds, then the two representations of 
the amplitude $\Ac$ are equivalent.

The final step is to show that Eq.~\eqref{eq:IntegralRK} is consistent with the reality of both $\Rc$ and $\Kc_{\df}$.
This result is not manifest, as $\Uc$ is complex.
Its imaginary part is readily found to be
\beq\label{eq:ImU}
\im \Uc_{\p'\p} = \frac{1}{3} \Big(\, 2 \bar\rho_{\p} \, \delta_{\p'\p} - \Cc_{\p'\p} \, \Big),
\eeq
where we have used the $\2\to\2$ unitarity relation for the inverse amplitude, 
\beq
\im \Fc_{\p}^{-1} = -\bar\rho_{\p} 
= - \rho_{\p} \Theta(\sigma_{\p} - 4m^2),
\label{eq:rhobar}
\eeq
which follows from Eq.~\eqref{eq:2to2UnitarityPW}, 
as well as the result $\Cc_{\p'\p} = \im \Gc_{\p'\p}$. 
To proceed we need the important results
\beq\label{eq:ImU_identity}
\int_{\k} \im \, \Uc_{\p'\k} \, \Kc_{\df}(\k,\p) = \int_{\k} \Kc_{\df}(\p',\k) \, \im \, \Uc_{\k\p} = 0.
\eeq
which are demonstrated in Appendix~\ref{sec:ImU_proof}. Essentially, the action of $\Cc_{\p'\p}$ on an object with the symmetry properties of Eq.~\eqref{eq:SymmK} yields a phase-space factor that cancels the first term of Eq.~\eqref{eq:ImU}. Combining these results, we find that $\Rc$ is real
if $\Kc_{\df}$ is,
\begin{align}
\im \Rc_{\p'\p} 
& =  \int_{\k'}\int_{\k} \Uc_{\p'\k'} \, \im \Kc_{\df}(\k',\k) \,\Uc_{\k\p}  \nn \\ 
\
& - \frac{1}{3} \int_{\k'}  \int_{\k} \Uc_{\p'\k'} \, \im \Kc_{\df}(\k',\k) \, \Rc_{\k\p} \nn \\
\
& - \frac{1}{3} \int_{\k'}  \int_{\k} \Uc_{\p'\k'} \,  \Kc_{\df}(\k',\k) \, \im \Rc_{\k\p} \nn \\
\
& = 0.
\end{align}
The inverse result can be shown similarly.
We conclude that the $B$-matrix and HS-BHS representations are equivalent.

\begin{figure*}[t!]
    \centering
    \includegraphics[ width=0.99\textwidth]{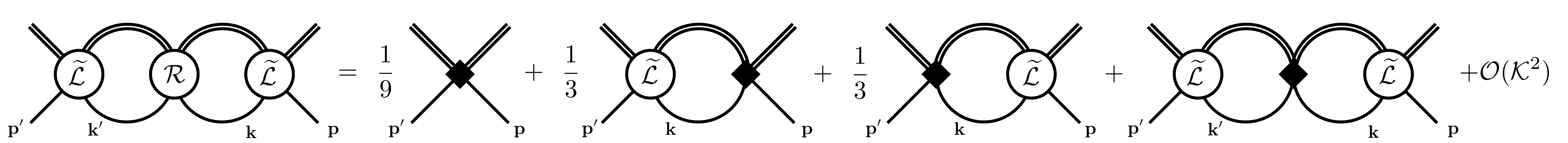}
    \caption{Diagrammatic representation of the leading order relation between $\Rc_{\p'\p}$ and $\Kc_{\df}(\p',\p)$, Eq.~\eqref{eq:compareRK}, where the black diamond represents $\Kc_{\df}(\p',\p)$.}
    \label{fig:diag_3to3_R_K_relations}
\end{figure*}

\begin{widetext}
The relationship between $\Rc_{\p'\p}$ and $\Kc_{\df}(\p',\p)$ in Eq.~\eqref{eq:IntegralRK} can be better understood if rewritten as
\begin{align}
\int_{\k'}\int_{\k} \wt{\Lc}_{\p'\k'} \,\Rc_{\k'\k} \, \wt{\Lc}_{\k\p} & = \int_{\k'} \int_{\k} \left( \frac{1}{3} \delta_{\p'\k'} + \wt{\Lc}_{\p'\k'} \, i\rho_{\k'} \right) \, \Kc_{\df}(\k',\k) \, \left( \frac{1}{3}\delta_{\k\p} + i\rho_{\k} \,\wt{\Lc}_{\k\p} \right) \nn \\
\
& -\frac{1}{3} \int_{\k'} \int_{\k} \int_{\k''} \left( \frac{1}{3} \delta_{\p'\k'} + \wt{\Lc}_{\p'\k'} \, i\rho_{\k'} \right) \, \Kc_{\df}(\k',\k) \, \Rc_{\k\k''} \, \wt{\Lc}_{\k''\p}.
\end{align}
We now assume that $\Kc_{\df}(\p',\p)$ is momentum-independent, $\Kc_{\df} =  \lambda$, with $\lambda$ a small constant.
This isotropic form is the leading contribution in an expansion
about threshold~\cite{Blanton:2019igq}.
Truncating the series solution of Eq.~\eqref{eq:IntegralRK} at leading order in $\Kc_{\df}$, we obtain
\begin{align}\label{eq:compareRK}
\int_{\k'}\int_{\k}\wt{\Lc}_{\p'\k'}\,\Rc_{\k'\k}\,\wt{\Lc}_{\k\p} & =  \frac{1}{9}\,\Kc_{\df}(\p'\p) 
\
 + \frac{1}{3} \int_{\k'} \wt{\Lc}_{\p'\k'} \, i\rho_{\k'} \, \Kc_{\df}(\k',\p)
\
 + \frac{1}{3} \int_{\k}  \Kc_{\df}(\p',\k) \, i\rho_{\k} \, \wt{\Lc}_{\k\p} \nn \\
\
& + \int_{\k'}\int_{\k} \wt{\Lc}_{\p'\k'} \, i\rho_{\k'} \, \Kc_{\df}(\k',\k) \, i\rho_{\k} \, \wt{\Lc}_{\k\p} + \Oc(\Kc_{\df}^{2}),
\end{align}
which is represented diagrammatically in Fig.~\ref{fig:diag_3to3_R_K_relations}. Since $\Kc_{\df}(\p',\p)$ represents three-body interactions such as contact interactions, the right hand side shows that there is a possibility that the interaction is not dressed by $\2\to\2$ rescatterings on either the initial or final state (or both). Contrarily, the $\Rc_{\p'\p}$ matrix is always dressed by $\2\to\2$ interactions in both the initial and final state. Thus, the $\Rc_{\p'\p}$ matrix represents a different organization of amplitudes. The factors $1/9$ and $1/3$ reflect the fact that $\Kc_{\df}(\p',\p)$ has a decomposition given by Eq.~\eqref{eq:SymmK}, which does not include summing over all spectator momenta. Thus, the factors are needed to remove the overcounting when we sum over all spectator momenta in the initial or final state. Finally, the left hand side has no $i\rho_{\p}$ factors, whereas the right hand side does. This is due to the differences in how the amplitudes are amputated. For the $B$-matrix, which is based on satisfying the unitarity relations, the amputation was made by removing the partial wave amplitudes via $\Ac_{\p'\p} = \Fc_{\p'} \, \wt{\Ac}_{\p'\p} \, \Fc_{\p}$
where the $\wt{\Ac}_{\p'\p}$ is the amputated partial wave amplitude. This was convenient as it simplified the unitarity relation (see Appendix~\ref{sec:unitarity} and Ref.~\cite{Jackura:2018xnx} for details). This amputation is not unique, as we can freely remove any real quantity from $\wt{\Ac}_{\p'\p}$, including $(i\rho_{\p})^2$. The HS-BHS equations involve an all-orders summation of amplitudes in an effective field theory, which includes loop integrals over four-momenta of intermediate states. When the two-particle loop integral is put on-shell, the $i\rho_{\p}$ factors naturally emerge. 

\end{widetext}

\section{Non-Relativistic Limit and Faddeev Equations}\label{sec:Faddeev}
In the non-relativistic limit, which is relevant for near threshold processes, we can investigate the relation between these representations to the Faddeev equations. 
If we assume that the three-body interactions are negligible compared to the two-body, then we can set the $\Kc_{\df}$-matrix (or equivalently the $R$-matrix by Eq.~\eqref{eq:IntegralRK}) to zero at leading order, leaving only the ladder rescattering solutions,
\beq
\Ac_{\p'\p} = \Dc_{\p'\p} + \Oc\left(\Kc_{\df}\right).
\eeq
The three-body amplitudes are dominated by exchanges between $\2\to\2$ processes, as in typical Faddeev-type approximations. In that case, as can be seen from Eqs. \eqref{eq:3to3_split} and \eqref{eq:Bmat_L}, the unsymmetrized scattering amplitude $\Mc$ becomes $\wt{\Lc}$, which gives
\beq\label{eq:AndrewFaddeev}
\Mc_{\p'\p} = \Fc_{\p}  \delta_{\p'\p} + \int_\k \Fc_{\p'} \Gc_{\p'\k} \Mc_{\k\p} \ .
\eeq
In the CMF, the non-relativistic limit of the $[\Gc_{\p'\p}]_{l'm'_l;lm_l}$ denominator in Eq. \eqref{eq:OPE} becomes
\begin{align}
& (P_\p - p')^2 - m^2 = (E_\p - \omega_{\p'})^2 - (\p+\p')^2 - m^2 \nn \\
& = 2m\left[ \Delta E - \frac{1}{m}\left(\p'^2 +\p^2 + \p \cdot \p' \right) \right] + \Oc(\p^4).
\end{align}
We approximated $\omega_\p = m + \p^2 / 2m + \Oc(\p^4 )$ and used the fact that, close to threshold, $E = 3 m +\Delta E$, where $\Delta E$ is a non-relativistic energy of three particles. Finally, we also neglected terms of the order $\p^4$,$\p'^4$, and $\p^2 \p'^2$. The factor $2\omega_\k$ in the integration measure of Eq. \eqref{eq:AndrewFaddeev} becomes $2m$. Putting everything together, and writing angular momentum indices explicitly, we obtain
\begin{widetext}
\begin{align}
[\Mc_{\p'\p}]_{\ell'm'_\ell;\ell m_\ell} &= \delta_{\p'\p} \delta_{\ell'm'_\ell;\ell m_\ell} [\Fc_{\p}]_{\ell m_\ell} \nn \\
&- \frac{4\pi}{(2\pi)^3 4m^2} \sum_{  \ell'', m_{\ell}'' } \int \diff^3 \k ~ [\Fc_{\p'}]_{\ell' m'_\ell} \left( \frac{k^{\star}}{q_{\p'}^{\star}} \right)^{\ell'}\!\!  \frac{  Y_{\ell' m_{\ell}'}({\bh{\k}}_{\p'}^{\star}) Y^*_{\ell'' m''_{\ell}}({\bh{\p}}_{\k}'^{\star}) }{\Delta E - \frac{1}{m}\left(\p'^2 +\k^2 + \k \cdot \p' \right) + i\epsilon}  \left( \frac{p'^{\star}}{q_{\k}^{\star}} \right)^{\ell} [\Mc_{\k\p} ]_{\ell''m''_\ell;\ell m_\ell} \ .
\end{align}
Following the conventions of Ref. \cite{LinElster:2007}, we define the non-relativistic scattering amplitude as ${}_{\textrm{NR}}^{\phantom{\prime}}{\braket{\textrm{out}|T|\textrm{in}}_{\textrm{NR}}} =  - 2\pi \,  \delta^{(4)}(P' - P) \, \Mc^{\textrm{NR}}$, 
which is different by a factor of $-(2\pi)^3$  to our definition in Eq. \eqref{eq:Smatrix}. Moreover, non-relativistic particle states are defined as $\ket{\p} = (2\pi)^{3/2} \, \sqrt{2\omega_{\p}} \, \ket{\p}_{\textrm{NR}}$, such that the momentum eigenstates are normalized as $_{\textrm{NR}}^{\phantom{\prime}}\!\braket{\p'|\p}_{\textrm{NR}} = \delta^{(3)}(\p' - \p)$. Thus, the relation between the non-relativistic and relativistic $\3\to\3$ amplitude is
\beq\label{eq:NR_amp_norm}
\Mc^{\textrm{NR}} = - \left( \frac{1}{(2\pi)^{3} \, 2m} \right)^{3} \Mc,
\eeq
where we have taken the nonrelativistic limit for the particle energies, \eg $\omega_{\p} = m + \Oc(\p^2)$. The $\2\to\2$ subprocesses contains an extra $\delta_{\p'\p} = (2\pi)^{3}\,2\omega_{\p} \, \delta^{3}(\p'-\p)$, which conserves spectator momenta, so the nonrelativistic $\2\to\2$ amplitude $\Fc_{\p}^{\textrm{NR}}$ is related to its relativistic
counterpart by an equation similar to Eq.~\eqref{eq:NR_amp_norm}, except that the conversion factor is squared instead of cubed. Therefore, after inclusion of spherical harmonics to recover full amplitudes, we arrive at
\beq
\label{eq:LudwigJackura}
\Mc_{\p'\p}^{\textrm{NR}}(\q_{\p'},\q_{\p}) = \delta^{3}(\p' - \p) \, \Fc_{\p}^{\textrm{NR}}(\q_{\p'},\q_{\p}) + \int \diff^{3}\k \, \Fc_{\p'}^{\textrm{NR}}(\q_{\p'},\k) \frac{1}{\Delta E - \frac{1}{m}(\p'^2 + \k^2 + \k \cdot \p') + i\epsilon} \, \Mc_{\k\p}^{\textrm{NR}}(\p',\q_{\p}) \ ,
\eeq
where
\begin{align}
\Mc_{\p'\p}^{\textrm{NR}}(\q_{\p'},\q_{\p}) &= 4\pi \sum_{ \substack{ \ell',m_{\ell}' \\ \ell,m_{\ell}} } Y_{\ell' m_\ell'}^{*}({\bh{\q}}_{\p'}) [\Mc^{\textrm{NR}}_{\p'\p}]_{\ell'm'_\ell;\ell m_\ell} Y_{\ell m_\ell}({\bh{\q}}_{\p})
\end{align}
and similarly for $\Fc_{\p'}^{\textrm{NR}}(\q_{\p'},\q_{\p})$. The barrier factors have been implicitly absorbed in $\Mc_{\p'\p}^{\textrm{NR}}$ and $\Fc_{\p}^{\textrm{NR}}$.
\end{widetext}

Equation~\eqref{eq:LudwigJackura} is equivalent to
the Faddeev equations, as can be seen, for example, by
comparing to Eq. (11) of Ref. \cite{Elster:2008hn}. There, a symmetrized two-body matrix $t_s$ plays a role of $\Fc_{\p}^{\textrm{NR}}$ and $T$ matrix plays a role of $\Mc_{\p'\p}^{\textrm{NR}}$. Note that both equations are not symmetrized with respect to interchanges of three interacting bosons as in Eq. \eqref{eq:SymmAmp}.

\section{Conclusions}\label{sec:Conclusion}
We have shown that the relativistic on-shell representation of
 the $\3\to\3$ scattering amplitude of Hansen and Sharpe~\cite{Hansen:2015zga} and
Brice\~no, Hansen, and Sharpe~\cite{Briceno:2017tce}, and the $B$-matrix representation presented by Mai et al. \cite{Mai:2017bge} and Jackura et al. \cite{Jackura:2018xnx} are equivalent, 
and their physical content is identical. The results of the present work are consistent with the conclusions of Ref.~\cite{Briceno:2019muc} that the HS-BHS approach is unitary. The difference in these representations is how the formalism incorporates rescattering effects. In the $B$-matrix representation, the $\3\to\3$ amplitude is always dressed by $\2\to\2$ rescatterings in both the initial and final states, as shown by Eq.~\eqref{eq:Bmat_convert}. Contrarily, the representation by Brice\~no, Hansen, and Sharpe allows the possibility of no initial/ final state rescatterings. It was shown in Section~\ref{sec:Relation_RK} that the differences between these rescattering functions manifest themselves as differences in the real part of the on-shell equations, giving the integral equation~\eqref{eq:IntegralRK}. The non-relativistic limit of both formalisms reproduced the Faddeev equations, providing a consistency check to well known low-energy approaches. 
As was discussed in Ref.~\cite{Jackura:2018xnx}, the $B$-matrix representations of Refs.~\cite{Mai:2017vot} and \cite{Jackura:2018xnx} 
differ only in the real part as a result of the latter
approach using a cut-off on the integration range that eliminated unphysical modes.

All of the proposed formalisms require regulation of the high-energy modes in order to arrive at a convergent solution to the integral equations. Regulating the divergent behavior introduces additional cutoff dependence in the equations.
Physical quantities must, however, be cutoff-independent, and this is achieved by introducing cutoff dependence into the real, $K$-matrix-like quantities in the formalisms (i.e.
$\Rc$ and $\Kc_{\df}$). For example, as was discussed in Ref.~\cite{Jackura:2018xnx}, the $B$-matrix representations of Ref.~\cite{Mai:2017vot} and \cite{Jackura:2018xnx} 
differ only in their real parts,
as a result of the latter using
a cutoff in the integration range which eliminated unphysical modes. 

It remains to be seen if the quantization conditions corresponding to the different formalisms are also identical. 
Naively, one might assume that, since the infinite volume equations are identical, the quantization conditions must also be, at least up to exponentially suppressed corrections. However, the details of transitioning from infinite to finite volume, \eg, the handling of angular momentum mixing, are nontrivial and have not yet been
worked out. This is an interesting area of study and must be completed to ensure consistency.

An interesting direction for future studies is comparing numerical results from each representation. Although equivalent, parametrizations using the $R$-matrix of Ref.~\cite{Jackura:2018xnx} or the $K$-matrix of Ref.~\cite{Hansen:2015zga} may turn out to be advantageous for
particular numerical analyses.

\begin{acknowledgments}
We thank Ra\'ul Brice\~no and Maxwell Hansen for many useful discussions.
This work was supported by
the U.S.~Department of Energy under grants
No.~DE-SC0011637 (SRS),
No.~DE-AC05-06OR23177, 
and No.~DE-FG02-87ER40365, 
U.S.~National Science Foundation under award number
PHY-1415459, 
PAPIIT-DGAPA (UNAM, Mexico) grant No.~IA101819, 
CONACYT (Mexico) grants No.~251817 and No.~A1-S-21389. 
VM acknowledges support from Comunidad Aut\'onoma de Madrid through 
Programa de Atracci\'on de Talento Investigador 2018 (Modalidad 1). The work of SRS was partly supported by the International Research Unit of Advanced Future Studies at Kyoto University.
\end{acknowledgments}

\appendix

\section{Unitarity Relations}\label{sec:unitarity}
Unitarity of the $S$-matrix constrains the imaginary part of on-shell scattering amplitudes. Given the unitarity constraints, one can construct an on-shell representation for scattering amplitudes in terms of real quantities and kinematic functions. We present here a brief summary of the unitarity relations for identical particles. The relations for distinguishable particles have been discussed in detail in Ref.~\cite{Jackura:2018xnx}. Elastic three-particle scattering satisfies the unitarity relation
\beq
2 \im \Mc = \frac{1}{3!}\prod_{j=1}^{3} \int_{\k_j} (2\pi)^{4} \delta^{(4)} \Bigg(\sum_{j=1}^{3}k_j -  P \Bigg) \Mc^{*} \, \Mc,
\eeq
where the integration is over the on-shell intermediate state momenta. Writing $\Mc$ in terms of the unsymmeterized amplitudes,  Eq.~\eqref{eq:SymmAmp}, and separating the disconnected $\2\to\2$ amplitude from the connected $\3\to\3$ amplitude via Eq.~\eqref{eq:3to3_split}, we arrive at two unitarity equations. The first is the well-known $\2\to\2$ unitarity relation in angular momentum space,
\beq\label{eq:2to2UnitarityPW}
\im \Fc_{\p} = \Fc_{\p}^{\dag} \, \bar\rho_{\p} \, \Fc_{\p},
\eeq
where $\bar\rho$ is the two-body phase space defined in Eq.~\eqref{eq:rhobar}.
Equation~\eqref{eq:2to2UnitarityPW} admits the on-shell $K$-matrix representation for $\Fc_{\p}$,
\begin{align}\label{eq:2to2_Kmat}
\Fc_{\p} & = \Kc_{\p} + \Kc_{\p} \, i\bar\rho_{\p} \, \Fc_{\p} \nn \\
\
& = \left[ 1 - \Kc_{\p} \,i\bar\rho_{\p} \right]^{-1} \Kc_{\p},
\end{align}
where $[ \, \Kc_{\p} \,]_{\ell' m_{\ell}' ; \ell m_{\ell}} = \delta_{\ell' \ell} \delta_{m_{\ell}' m_{\ell}} \Kc_{\ell}(\sigma_{\p})$ is the $\2\to\2$ $K$-matrix, which is a real function of $\sigma_{\p}$ in the elastic kinematic region, and diagonal in angular momenta space. Since the phase space factor contains the kinematic information of two on-shell propagating particles, the $K$-matrix represents all the dynamical content of the two-particle system, \eg such as the short range forces between pions in elastic $\pi\pi$ scattering. This can in principle include virtual exchanges leading to left hand cuts or higher multiparticle thresholds, \eg four particle production, which do not give singular contributions in the elastic domain.
Since $\2\to\2$ amplitudes are diagonal in angular momentum space, Eq.~\eqref{eq:2to2_Kmat} reduces to a simple algebraic relation. It is straightforward to verify that Eq.~\eqref{eq:2to2_Kmat} satisfies Eq.~\eqref{eq:2to2UnitarityPW}.

\begin{figure}[t!]
    \centering
    \includegraphics[ width=0.48\textwidth]{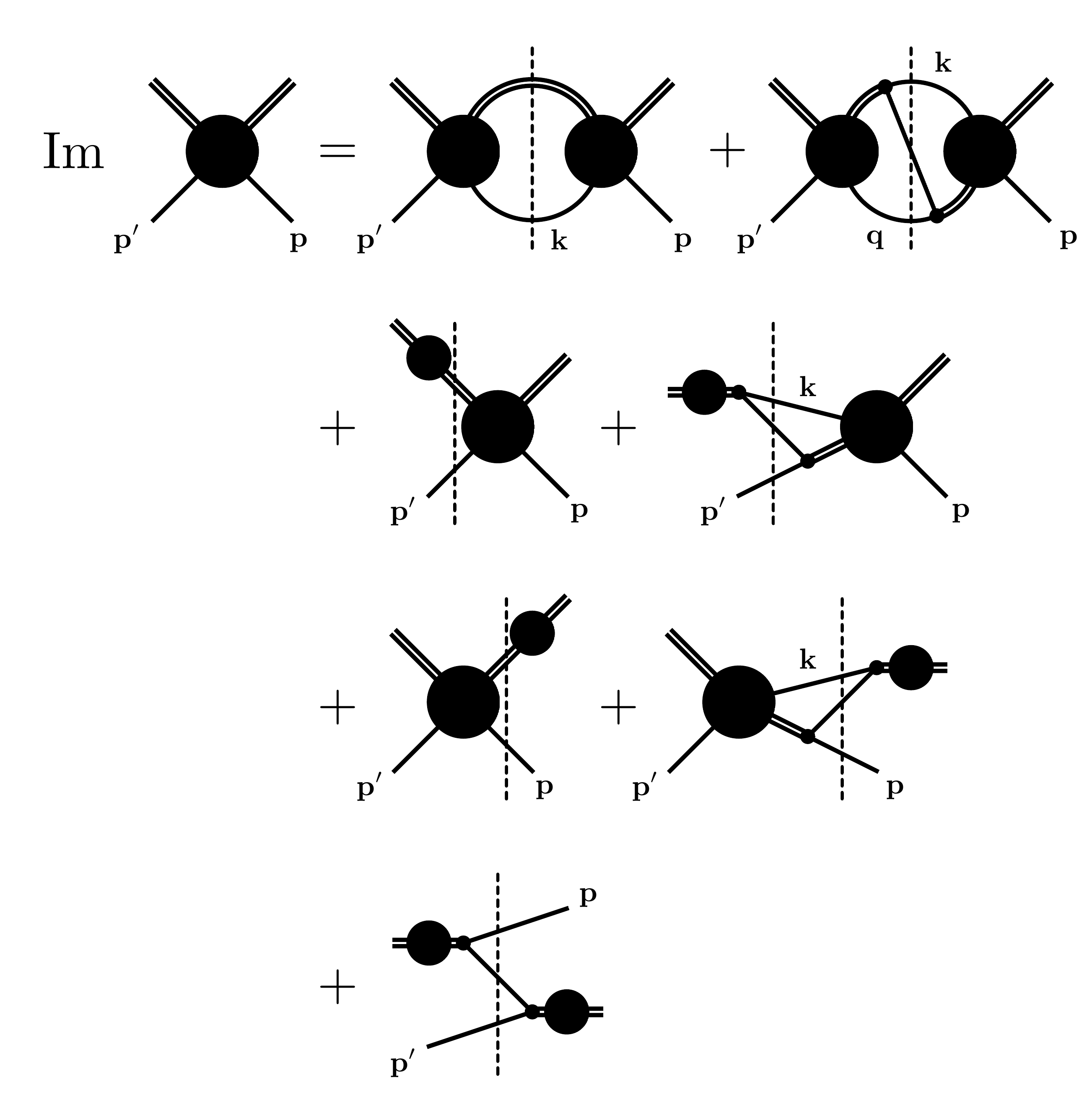}
    \caption{Diagrammatic representation for the $\3\to\3$ unitarity relation for the amplitude $\Ac_{\ell' m_{\ell}' ; \ell m_{\ell}}(\p',s,\p)$. Closed loops yield three-dimensional integrations over the labeled spectator momentum, and the dashed vertical lines represent placing all three intermediate state particles on their mass-shell. Momentum flow is from right to left, as before, and each amplitude on the left of the dashed line is hermitian conjugated.}
    \label{fig:diag_3to3_unitarity}
\end{figure}

The second unitarity relation is for the connected $\3\to\3$ amplitude, which in the $p\ell m_{\ell}$-basis is
\begin{align}\label{eq:3to3unitarity}
\im \Ac_{\p'\p} & = \int_{\k} \Ac_{\p'\k}^{\dag} \, \bar\rho_{\k} \, \Ac_{\k\p} + \int_{\k'}\int_{\k} \Ac_{\p'\k'}^{\dag} \,  \Cc_{\k'\k} \,  \Ac_{\k\p} \nn \\
\
& + \Fc_{\p'}^{\dag} \, \bar\rho_{\p'} \, \Ac_{\p'\p} + \int_{\k} \Fc_{\p'}^{\dag} \, \Cc_{\p'\k} \, \Ac_{\k \p} \nn \\
\
& +  \Ac_{\p'\p}^{\dag} \, \bar\rho_{\p} \, \Fc_{\p} + \int_{\k} \Ac_{\p'\k}^{\dag} \,  \Cc_{\k\p} \, \Fc_{\p} \nn \\
\
& + \Fc_{\p'}^{\dag} \, \Cc_{\p'\p} \, \Fc_{\p},
\end{align}
where $\Cc_{\p'\p}$ is the recoupling coefficient between a pair in one state to a different pair in the same state, \eg, from an angular momentum coupling (12)3 to (23)1, which is defined as the imaginary part of the amputated OPE amplitude, Eq.~\eqref{eq:OPE},
\begin{align}
\label{eq:recoupling}
\left[ \, \Cc_{\p'\p} \, \right]_{\ell' m_{\ell}' ; \ell m_{\ell}} & \equiv \im \left[ \,\Gc_{\p'\p} \, \right]_{\ell' m_{\ell}' ; \ell m_{\ell}}  \nn \\
\
& = \pi \, \delta\left( (P_{\p} - p' )^2 - m^2 \right) \nn \\
& \times   4\pi \, Y_{\ell' m_{\ell}'}({\bh{\p}}_{\p'}^{\star}) Y^*_{\ell m_{\ell}}({\bh{\p}}_{\p}'^{\star}) .
\end{align}
The recoupling coefficients are an additional feature of three-body scattering that can be seen in Fig.~\ref{fig:diag_3to3_unitarity} when a diagram with a crossed exchange in the intermediate state is cut.  Diagrams that are cut where no exchange occurs gives rise to the conventional two-body phase space.
One may be concerned that the complexity of spherical harmonics is not taken into account. The phases in the unitarity relation cancel since the intermediate state sums over all possibilities. To avoid this bookkeeping during intermediate calculations, one can use real spherical harmonics, which have the same completeness and orthonormality relations as the usual ones, to formally manipulate the expressions. Since the final results do not depend on the choice of harmonics, we are guaranteed the validity of the unitarity relations and the solutions.

Equation~\eqref{eq:3to3unitarity} admits the on-shell representation given by Eq.~\eqref{eq:Bmatrix}, which we now verify. We find the following demonstration more direct than the one presented in Ref.~\cite{Jackura:2018xnx}. First, let us introduce amplitudes which have the final state $\2\to\2$ amplitudes amputated, \ie, $\Ac_{\p'\p} = \Fc_{\p'} \, \wt{\Ac}_{\p'\p} \, \Fc_{\p}$. Equation~\eqref{eq:3to3unitarity} then simplifies to
\begin{align}\label{eq:3to3unitarity_amputated}
\im \wt{\Ac}_{\p'\p} & = \int_{\k} \wt{\Ac}_{\p'\k}^{\dag} \, \Fc_{\k}^{\dag} \, \bar\rho_{\k} \, \Fc_{\k} \, \wt{\Ac}_{\k\p} \nn \\
\
& + \int_{\k'}\int_{\k} \wt{\Ac}_{\p'\k'}^{\dag} \, \Fc_{\k'}^{\dag} \, \Cc_{\k'\k} \, \Fc_{\k} \, \wt{\Ac}_{\k\p} \nn \\
\
& + \int_{\k}  \Cc_{\p'\k} \, \Fc_{\k} \, \wt{\Ac}_{\k \p}  + \int_{\k} \wt{\Ac}_{\p'\k}^{\dag} \, \Fc_{\k}^{\dag} \,  \Cc_{\k\p}  \nn \\
\
& +  \Cc_{\p'\p},
\end{align}
and the corresponding amputated $B$-matrix representation is
\begin{align}\label{eq:Bmatrix_app}
\wt{\Ac}_{\p'\p} & = \Bc_{\p'\p} + \int_{\k} \Bc_{\p'\k} \, \Fc_{\k} \, \wt{\Ac}_{\k\p} \nn \\
\
& = \int_{\k} \Bc_{\p'\k} \left( \delta_{\k\p} + \Fc_{\k} \, \wt{\Ac}_{\k\p} \right),
\end{align}
where we remind the reader that $\Bc_{\p'\p} = \Gc_{\p'\p} + \Rc_{\p'\p}$, where $\Gc_{\p'\p}$ is given in Eq.~\eqref{eq:OPE} and $\Rc_{\p'\p}$ is a real function that contains the unconstrained three-body dynamics.
It is straightforward to verify that Eq.~\eqref{eq:Bmatrix_app} satisfies Eq.~\eqref{eq:3to3unitarity_amputated} directly by taking the difference between the amplitude and its Hermitian conjugate. Note that if the matrix elements of $\Ac_{\p'\p}$ are $\Ac_{\ell' m_{\ell}' ; \ell m_{\ell}}(\p',s,\p)$, then the Hermitian conjugate, $\Ac_{\p'\p}^{\dag}$, has elements $\Ac_{\ell m_{\ell} ; \ell' m_{\ell}'}^{*}(\p,s,\p')$, since it acts on both the angular momentum space and the spectator space. The Hermitian analytic properties \cite{Eden:1966dnq} of amplitudes then state $\Ac_{\ell m_{\ell} ; \ell' m_{\ell}'}^{*}(\p,s,\p') = \Ac_{\ell' m_{\ell}' ; \ell m_{\ell}}^{*}(\p',s,\p)$, so that $\Ac_{\p'\p} - \Ac_{\p'\p}^{\dag} = \Ac_{\p'\p} - \Ac_{\p'\p}^{*} = 2i \, \im \Ac_{\p'\p}$. 

We begin by rewriting the difference by adding and subtracting a judiciously chosen term, leading to
\begin{align}\label{eq:3to3Proof_1}
\wt{\Ac}_{\p'\p} - \wt{\Ac}_{\p'\p}^{\dag} & = \int_{\k} \wt{\Ac}_{\p'\k}^{\dag} \Big( \Fc_{\k} - \Fc_{\k}^{\dag} \Big) \wt{\Ac}_{\k\p} \nn \\
\
& + \int_{\k} \left( \delta_{\p'\k} + \wt{\Ac}_{\p'\k}^{\dag} \Fc_{\k}^{\dag}\right)\wt{\Ac}_{\k\p} \nn \\
\
& - \int_{\k} \wt{\Ac}_{\p'\k}^{\dag}\left( \delta_{\k \p} + \Fc_{\k} \wt{\Ac}_{\k \p} \right) .
\end{align}
Next we insert Eq.~\eqref{eq:Bmatrix_app} into $\wt{\Ac}_{\k\p}$ on the second line of Eq.~\eqref{eq:3to3Proof_1}, and its Hermitian conjugate,
\begin{align}
\wt{\Ac}_{\p'\p}^{\dag} & = \Bc_{\p'\p}^{\dag} + \int_{\k} \wt{\Ac}_{\p'\k}^{\dag} \, \Fc_{\k}^{\dag} \, \Bc_{\k\p}^{\dag} \nn \\
\
& = \int_{\k} \left( \delta_{\p'\k} + \wt{\Ac}_{\p'\k}^{\dag} \, \Fc_{\k}^{\dag}  \right) \Bc_{\k\p}^{\dag},
\end{align}
into the $\wt{\Ac}_{\p'\k}^{\dag}$ on the third line of Eq.~\eqref{eq:3to3Proof_1}. This gives
\begin{align}\label{eq:3to3Proof_2}
\wt{\Ac}_{\p'\p} - \wt{\Ac}_{\p\p'}^{\dag} & = \int_{\k} \wt{\Ac}_{\p'\k}^{\dag} \Big( \Fc_{\k} - \Fc_{\k}^{\dag} \Big) \wt{\Ac}_{\k\p} \nn \\
\
& + \int_{\k'}\int_{\k} \left(\, \delta_{\p'\k'} + \wt{\Ac}_{\p'\k}^{\dag} \, \Fc_{\k'}^{\dag} \,\right) \nn \\
\
& \times \left( \, \Bc_{\k'\k} - \Bc_{\k'\k}^{\dag} \, \right) \nn \\
\
& \times \left( \, \delta_{\k\p} + \Fc_{\k} \, \wt{\Ac}_{\k\p} \, \right),
\end{align}
which can then be simplified using $\Fc_{\p} - \Fc_{\p}^{\dag} = 2i \, \im \Fc_{\p}$ and Eq.~\eqref{eq:2to2UnitarityPW}, as well as the result that $\Bc_{\p'\p} - \Bc_{\p'\p}^{\dag} = 2i \, \im \Gc_{\p'\p}$ since $\Rc_{\p'\p}$ is real. Then, since the recoupling coefficients are $\Cc_{\p'\p} = \im \Gc_{\p'\p}$, we arrive at Eq.~\eqref{eq:3to3unitarity_amputated}, thus proving that the $B$-matrix representation satisfies the unitary condition.

\section{Expressing the B-Matrix in terms of the Ladder}\label{sec:Reexpress}
In this appendix, we show how the $B$-matrix representation can be expressed in terms of the full OPE ladder summation and a remaining piece containing genuine three-body interactions (see also Ref.~\cite{Mikhasenko:2019vhk}).
The $B$-matrix representation for the full amplitude is given in Eq.~\eqref{eq:Bmatrix}.
In the limit that the scattering is dominated by $\2\to\2$ interactions, and three-body interactions are negligible ($\Rc_{\p'\p} \to 0$), then the system is controlled by successive particle exchanges between the $\2\to\2$ amplitudes. We defined this process as the ladder amplitude, $\Dc_{\p'\p}$, which satisfies Eq.~\eqref{eq:OPE_ladder}. 
We now want to remove the ladder solution from the general three-body system. In the same vein as HS-BHS, we define the divergence-free amplitude, $\Ac_{\df,\p'\p} \equiv \Ac_{\p'\p} - \Dc_{\p'\p}$, which the $\3\to\3$ amplitude  free from the ladder diagram and its singularities. We can then separate the ladder solution from the $B$-matrix representation and are left with an equation for $\Ac_{\df,\p'\p}$,
\begin{align}\label{eq:Bmat_DF}
\Ac_{\df,\p'\p}  & = \int_{\k} \Fc_{\p'}\, \Rc_{\p'\k} \, \left( \Fc_{\k} \delta_{\k\p} + \Dc_{\k\p} \right) \nn \\
& + \int_{\k}\Fc_{\p'}\, \left( \Rc_{\p'\k} + \Gc_{\p'\k} \right) \,\Ac_{\df,\k\p} .
\end{align}
Now define the $\2\to\2$ rescattering function Eq.~\eqref{eq:Bmat_L}, and amputate the end caps from the divergent free amplitude,
\beq\label{eq:amputate_app}
\Ac_{\df,\p'\p} = \int_{\k'}\int_{\k} \wt{\Lc}_{\p\k} \, \wt{\Tc}_{\k'\k} \, \wt{\Lc}_{\k\p}.
\eeq
Substituting Eq.~\eqref{eq:amputate_app} into Eq.~\eqref{eq:Bmat_DF} and collecting terms, we arrive at
\begin{align}
\int_{\k'}\int_{\k} & \left[ \, \delta_{\p'\k'} - \Fc_{\p'} \Gc_{\p'\k'} \, \right] \, \wt{\Lc}_{\k'\k} \, \wt{\Tc}_{\k \p} \nn \\
\
& = \Fc_{\p'} \, \Rc_{\p'\p} + \int_{\k'}\int_{\k} \Fc_{\p'} \, \Rc_{\p'\k'} \, \wt{\Lc}_{\k'\k} \, \wt{\Tc}_{\k\p},
\end{align}
where we have removed the right-most rescattering function, and collected all terms with $\Rc_{\p'\p}$ on the right hand side. Finally, the combination on the left hand side simplifies to
\begin{align}
\int_{\k'} & \left[ \, \delta_{\p'\k'} - \Fc_{\p'} \Gc_{\p'\k'} \, \right] \, \wt{\Lc}_{\k'\k}  = \Fc_{\p'} \delta_{\p'\k} \nn \\
\
& + \Bigg( \Dc_{\p'\k} - \Fc_{\p'} \, \Gc_{\p'\k} \, \Fc_{\k} - \int_{\k'} \Fc_{\p'} \, \Gc_{\p'\k'} \, \Dc_{\k'\k}\Bigg),
\end{align}
where the term inside the parenthesis is zero from Eq.~\eqref{eq:OPE_ladder}. Factorizing the final $\2\to\2$ amplitude from the left hand side, we arrive at the resummed $\3\to\3$ amplitude,
\beq\label{eq:Amp_rex_app}
\Ac_{\p'\p} = \Dc_{\p'\p} + \int_{\k'}\int_{\k} \wt{\Lc}_{\p'\k'} \, \wt{\Tc}_{\k'\k} \, \wt{\Lc}_{\k\p},
\eeq
with the new amputated amplitude satisfying
\beq\label{eq:Tamp_app}
\wt{\Tc}_{\p'\p} = \Rc_{\p'\p} + \int_{\k'}\int_{\k} \Rc_{\p'\k'} \, \wt{\Lc}_{\k'\k} \, \wt{\Tc}_{\k\p}.
\eeq

Equation \eqref{eq:Amp_rex_app}, along with Eqs.~\eqref{eq:OPE_ladder} and \eqref{eq:Tamp_app}, is an alternative on-shell representation for the $\3\to\3$ scattering amplitude that satisfies unitarity. We now proceed with similar manipulations on the unitarity relation, Eq.~\eqref{eq:3to3unitarity}, allowing one to derive Eq.~\eqref{eq:Tamp_app} directly from unitarity.

It is clear from the demonstration in Appendix~\ref{sec:unitarity} that $\Dc_{\p'\p}$ satisfies the same unitarity relation as Eq.~\eqref{eq:3to3unitarity}. Therefore, the unitarity relation for the divergence-free amplitude states
\begin{align}
\im \Ac_{\df,\p'\p} & = \int_{\k} \int_{\q} \Ac_{\df,\p'\q}^{\dag} \, \Big( \, \bar{\rho}_{\q} \, \delta_{\q\k}  + \Cc_{\q\k} \, \Big) \, \Ac_{\df,\k\p}  \nn \\
\
& + \int_{\k} \int_{\q} \Ac_{\df,\p'\q}^{\dag} \, \Big( \, \bar{\rho}_{\q} \, \delta_{\q\k}  + \Cc_{\q\k} \, \Big) \, 
\wt{\Lc}_{\k\p} \nn \\
\
& + \int_{\k} \int_{\q} \wt{\Lc}_{\p'\q}^{\dag} \, \Big( \, \bar{\rho}_{\q} \, \delta_{\q\k}  + \Cc_{\q\k} \, \Big) \, \Ac_{\df,\k\p}  .
\end{align}
 Since $\Fc_{\p}$ obeys the $\2\to\2$ unitarity relation, Eq.~\eqref{eq:2to2UnitarityPW}, and $\Dc_{\p'\p}$ satisfies Eq.~\eqref{eq:3to3unitarity}, we can see that the rescattering function $\wt{\Lc}_{\p'\p}$ satisfies the relation
\begin{align}
\label{eq:rescattering_im_part}
\im \wt{\Lc}_{\p'\p} & = \int_{\k} \wt{\Lc}_{\p'\k}^{\dag} \, \bar{\rho}_{\k} \, \wt{\Lc}_{\k\p} \nn \\
\
& + \int_{\k'}\int_{\k} \wt{\Lc}_{\p'\k'}^{\dag} \, \Cc_{\k'\k} \, \wt{\Lc}_{\k \p}.
\end{align}
The amputated divergence-free amplitude can be defined as in Eq.~\eqref{eq:amputate_app}, so that the unitarity relation becomes
\begin{align}\label{eq:3to3Unit_alt}
\im \wt{\Tc}_{\p'\p} & = \int_{\k'}\int_{\k} \int_{\q} \wt{\Tc}_{\p'\k'}^{\dag}  \, \wt{\Lc}_{\k'\q}^{\dag} \, \bar{\rho}_{\q} \, \wt{\Lc}_{\q\k} \,  \wt{\Tc}_{\k\p} \nn \\
\
& + \int_{\k'}\int_{\k} \int_{\q'} \int_{\q} \wt{\Tc}_{\p'\k'}^{\dag} \,  \wt{\Lc}_{\k'\q'}^{\dag} \, \Cc_{\q'\q} \, \wt{\Lc}_{\q\k}  \, \wt{\Tc}_{\k\p} \nn \\
\
& = \int_{\k'} \int_{\k} \wt{\Tc}_{\p'\k'}^{\dag} \, \im \wt{\Lc}_{\k'\k} \, \wt{\Tc}_{\k\p}.
\end{align}
Using similar manipulations as in Appendix~\ref{sec:unitarity}, it is straightforward to verify that Eq.~\eqref{eq:Tamp_app} satisfies the unitarity relation Eq.~\eqref{eq:3to3Unit_alt}.

\begin{widetext}
\section{Proof of Eq.~\eqref{eq:ImU_identity}}\label{sec:ImU_proof}
In Sec.~\ref{sec:Relation_RK}, we showed that the $R$-matrix and three-body $K$-matrix are related by an integral equation, Eq.~\eqref{eq:IntegralRK}. Proving the reality of the Eq.~\eqref{eq:IntegralRK} relied on the claim  Eq.~\eqref{eq:ImU_identity}, which we now prove.

From the definition, Eq.~\eqref{eq:Ufactor}, we find that
\begin{align}\label{eq:app_U_proof}
3\sum_{\ell'',m_{\ell}''}\int_{\k} \im\, \left[ \, \Uc_{\p'\k} \, \right]_{\ell' m_{\ell}' ; \ell'' m_{\ell}''} & \, \left[ \, \Kc_{\df}(\k,\p) \, \right]_{\ell'' m_{\ell}'' ; \ell m_{\ell}} \nn \\
\
& = \sum_{\ell'',m_{\ell}''} \int_{\k} \Big[ \, \, 2\bar\rho_{\p'} \delta_{\p'\k} - \Cc_{\p'\k} \, \Big]_{\ell' m_{\ell}' ; \ell m_{\ell}}\left[ \,  \Kc_{\df}(\k,\p) \, \right]_{\ell'' m_{\ell}'' ; \ell m_{\ell}} \nn \\
& =\sum_{\ell'',m_{\ell}''} \delta_{\ell' \ell''} \delta_{m_{\ell}'m_{\ell}''} \, 2\bar\rho_{\p'} \, \Kc_{\df,\ell'' m_{\ell}'' ; \ell m_{\ell}}(\p',\p) \nn \\
\
& - \sum_{\ell'',m_{\ell}''} \int_{\k} \pi \delta\left( \left( P - k  - p' \right)^2  - m^2\right) 4\pi Y_{\ell' m_{\ell}'}({\bh{\k}}^{\star}) Y^*_{\ell'' m_{\ell}''}({\bh{\p}}'^{\star}) \, \Kc_{\df,\ell'' m_{\ell}'' ; \ell m_{\ell}}(\k,\p),
\end{align}
where in the second term, Eq. \eqref{eq:recoupling} was used. We leave the first term as is, and focus on the second term. According to Ref.~\cite{Hansen:2015zga}, $\Kc_{\df}(\k,\p)$ is defined as a symmetric object after acting with spherical harmonics of the pair orientations on $\Kc_{\df}(\k,\p)$, and summing over all angular momenta. We use this property to combine the product of the final spherical harmonic $Y_{\ell'' m_{\ell}''}({\bh{\p}}'^{\star})$ and $\Kc_{\df}(\k,\p)$, and then switch the role of $\p'$ and $\k$, finally expanding in spherical harmonics of ${\bh{\k}}^\star$. This allows us to write
\beq
\sum_{\ell'',m_{\ell}''} Y^*_{\ell'' m_{\ell}''}({\bh{\p}}'^{\star}) \, \Kc_{\df,\ell'' m_{\ell}'' ; \ell m_{\ell}}(\k,\p) = \sum_{\ell'',m_{\ell}''} Y^*_{\ell'' m_{\ell}''}({\bh{\k}}^{\star}) \, \Kc_{\df,\ell'' m_{\ell}'' ; \ell m_{\ell}}(\p',\p),
\eeq
Now, $\Kc_{\df}(\p',\p)$ is independent of $\k$, thus we can perform the integrations
\begin{align}
\sum_{\ell'', m_{\ell}''}\int_{\k} \pi \delta & \left( \left( P - k - p' \right)^2 - m^2 \right)  \, 4\pi Y_{\ell' m_{\ell}'}({\bh{\k}}^{\star}) Y^*_{\ell'' m_{\ell}''}({\bh{\k}}^{\star}) \, \Kc_{\df,\ell'' m_{\ell}'' ; \ell m_{\ell}}(\p',\p) \nn \\
\
& = \frac{1}{4\pi} \int_{0}^{\infty} \diff k^{\star} \frac{k^{\star\,2}}{4\omega_{\k^{\star}}^2} \, \delta( \omega_{\k^{\star}} - E_{\p'}^{\star} / 2 ) \sum_{\ell'', m_{\ell}''} \int \diff {\bh{\k}}^{\star} \, Y_{\ell' m_{\ell}'}({\bh{\k}}^{\star}) Y^*_{\ell'' m_{\ell}''}({\bh{\k}}^{\star}) \, \Kc_{\df,\ell'' m_{\ell}'' ; \ell m_{\ell}}(\p',\p) \nn \\
\
& = 2 \bar\rho_{\p'} \, \sum_{\ell'',m_{\ell}''} \delta_{\ell' \ell''} \delta_{m_{\ell}' m_{\ell}''} \, \Kc_{\df,\ell'' m_{\ell}'' ; \ell m_{\ell}}(\p',\p),
\end{align}
\end{widetext}
where we converted to spherical coordinates in the final pair rest frame, and used the composition properties of Dirac delta functions to convert to the on-shell energy $\omega_{\k^{\star}}$. Orthogonality properties of spherical harmonics allows the angular integration to be done, showing that the second term is identical to the first of Eq.~\eqref{eq:app_U_proof}. Thus, we conclude that
\beq
\int_{\k} \im \, \Uc_{\p'\k} \, \Kc_{\df}(\k,\p) = 0,
\eeq
as claimed. The relation, $\int_{\k} \Kc_{\df}(\p',\k) \,\im \, \Uc_{\k\p} = 0$, is verified in an identical manner.

\bibliographystyle{apsrev4-1}
\bibliography{3to3.bib}

\end{document}

%% file: shortcuts.tex
\renewcommand{\k}{\mathbf{k}}
\newcommand{\p}{\mathbf{p}}
\newcommand{\q}{\mathbf{q}}

\renewcommand{\P}{\mathbf{P}}

\newcommand{\0}{\mathbf{0}}

\newcommand{\2}{\mathbf{2}}
\newcommand{\3}{\mathbf{3}}

\newcommand{\Ac}{\mathcal{A}}
\newcommand{\Bc}{\mathcal{B}}
\newcommand{\Cc}{\mathcal{C}}
\newcommand{\Dc}{\mathcal{D}}
\newcommand{\Ec}{\mathcal{E}}
\newcommand{\Fc}{\mathcal{F}}
\newcommand{\Gc}{\mathcal{G}}

\newcommand{\Kc}{\mathcal{K}}
\newcommand{\Lc}{\mathcal{L}}
\newcommand{\Mc}{\mathcal{M}}

\newcommand{\Oc}{\mathcal{O}}

\newcommand{\Rc}{\mathcal{R}}
\newcommand{\Sc}{\mathcal{S}}
\newcommand{\Tc}{\mathcal{T}}
\newcommand{\Uc}{\mathcal{U}}

\DeclareMathOperator{\im}{Im}

\newcommand{\bs}[1]{\boldsymbol{ #1 }}

\newcommand{\wt}[1]{\widetilde{ #1 }}

\newcommand{\bh}[1]{\bf{\hat{ #1 }}}

\newcommand{\eg}{e.g.\xspace}

\newcommand{\ie}{i.e.\xspace}

\newcommand{\df}{\textrm{df}}
\newcommand{\nn}{\nonumber}
\newcommand{\diff}{\textrm{d}}

\newcommand{\beq}{\begin{equation}}
\newcommand{\eeq}{\end{equation}}

\newcommand{\kev}{\ensuremath{{\mathrm{\,ke\kern -0.1em V}}}\xspace}
\newcommand{\mev}{\ensuremath{{\mathrm{\,Me\kern -0.1em V}}}\xspace}
\newcommand{\gev}{\ensuremath{{\mathrm{\,Ge\kern -0.1em V}}}\xspace}
\newcommand{\tev}{\ensuremath{{\mathrm{\,Te\kern -0.1em V}}}\xspace}

\usepackage{mfirstuc} 
\newcommand{\addReviewer}[2]{
  \expandafter\newcommand\csname #1\endcsname[1]{{\bf \color{#2} \capitalisewords{#1}:\,##1}}
  \expandafter\newcommand\csname #1cor\endcsname[2]{{\color{#2} \capitalisewords{#1}:\,\st{##1}{\bf ##2}}}
  \expandafter\newcommand\csname #1color\endcsname{#2}
}

\usepackage{soul,color}
\definecolor{chromeyellow}{rgb}{1.0, 0.65, 0.0}
\definecolor{DodgeBlue}{rgb}{0.118, 0.565,1.000}
\definecolor{asparagus}{rgb}{0.53, 0.66, 0.42}
\definecolor{cardinal}{rgb}{0.77, 0.12, 0.23}
\definecolor{cadmiumgreen}{rgb}{0.0, 0.42, 0.24}
\definecolor{applegreen}{rgb}{0.55, 0.71, 0.0}

\addReviewer{andrew}{cadmiumgreen}
\addReviewer{sebastian}{cardinal}
\addReviewer{cesar}{red}
\addReviewer{misha}{applegreen}
\addReviewer{ale}{chromeyellow}